\documentclass[twocolumn,usenatbib]{mn2e}
\usepackage{psfig}
\title[SN~2004dj in NGC~2403]{The first year of SN~2004dj in NGC~2403\thanks{Based
on observations obtained at David Dunlap Observatory (Canada), F.L.Whipple Observatory
(USA), Konkoly Observatory and Szeged Observatory (Hungary)}}
\author[J. Vink\'o et al.]{ J. Vink\'o$^{1}$\thanks{E-mail:vinko@physx.u-szeged.hu},
K. Tak\'ats$^{1}$,  K. S\'arneczky$^1$, Gy. M. Szab\'o$^2$, Sz. M\'esz\'aros$^{1,5}$, 
\newauthor
R. Csorv\'asi$^1$,T. Szalai$^1$, A. G\'asp\'ar$^1$, A. P\'al$^3$, Sz. Csizmadia$^4$, A. K\'osp\'al$^4$, 
\newauthor
M. R\'acz$^4$, M. Kun$^4$, B. Cs\'ak$^{1,5}$, G. F\H {u}r\'esz$^{4,5}$, H. DeBond$^6$, J. Grunhut$^6$,
\newauthor
J. Thomson$^6$, S. Mochnacki$^6$, T. Koktay$^6$\\
$^1$ Dept. of Optics \& Quantum Electronics, University of Szeged, Hungary\\
$^2$ Dept. of Experimental Physics, University of Szeged, Hungary\\
$^3$ Dept. of Astronomy, ELTE University, Budapest, Hungary\\
$^4$ Konkoly Observatory of Hungarian Academy of Sciences, Budapest, Hungary\\
$^5$ Harvard-Smithsonian Center for Astrophysics, Cambridge MA, USA\\
$^6$ David Dunlap Observatory, University of Toronto, Richmond Hill ON, Canada
 }

\begin{document}

\date{Accepted . Received ; in original form }

\pagerange{\pageref{firstpage}--\pageref{lastpage}} \pubyear{2006}

\maketitle

\label{firstpage}

\begin{abstract}
New $BVRI$ photometry and optical spectroscopy of the Type IIp supernova 2004dj
in NGC~2403, obtained during the first year since discovery, are presented. 
The progenitor cluster, Sandage~96, is also detected on pre-explosion
frames.
The light curve indicates that the explosion occured about 30 days before discovery,
and the plateau phase lasted about $+110 \pm 20$ days after that. The  plateau-phase 
spectra have been modelled with the $SYNOW$ spectral synthesis code
using H, Na~I, Ti~II, Sc~II, Fe~II and Ba~II lines. The SN distance is inferred
from the Expanding Photosphere Method (EPM) and the Standard Candle Method (SCM)
applicable for SNe IIp. They resulted in distances that are consistent with each
other as well as earlier Cepheid- and Tully-Fisher distances.
The average distance, $D = 3.47 \pm 0.29$ Mpc is proposed for SN~2004dj and NGC~2403.
The nickel mass produced by the explosion is estimated
as $\sim 0.02 \pm 0.01 ~M_{\odot}$. The SED of the progenitor cluster
is reanalysed by fitting population synthesis models to our observed $BVRI$ data
supplemented by $U$ and $JKH$ magnitudes from the literature. The $\chi^2$-minimization
revealed a possible "young" solution with cluster age $T_{cl} = 8$ Myr, and an
"old" solution with $T_{cl} = 20$ - $30$ Myr. The "young" solution would imply
a progenitor mass $M > 20 ~M_{\odot}$, which is higher than the previously
detected progenitor masses for Type II SNe.  
\end{abstract}

\begin{keywords}
stars: evolution -- supernovae: individual (SN~2004dj) -- galaxies: individual (NGC~2403) 
\end{keywords}

\section{Introduction}

\begin{figure*}
\begin{center}
\leavevmode
\psfig{file=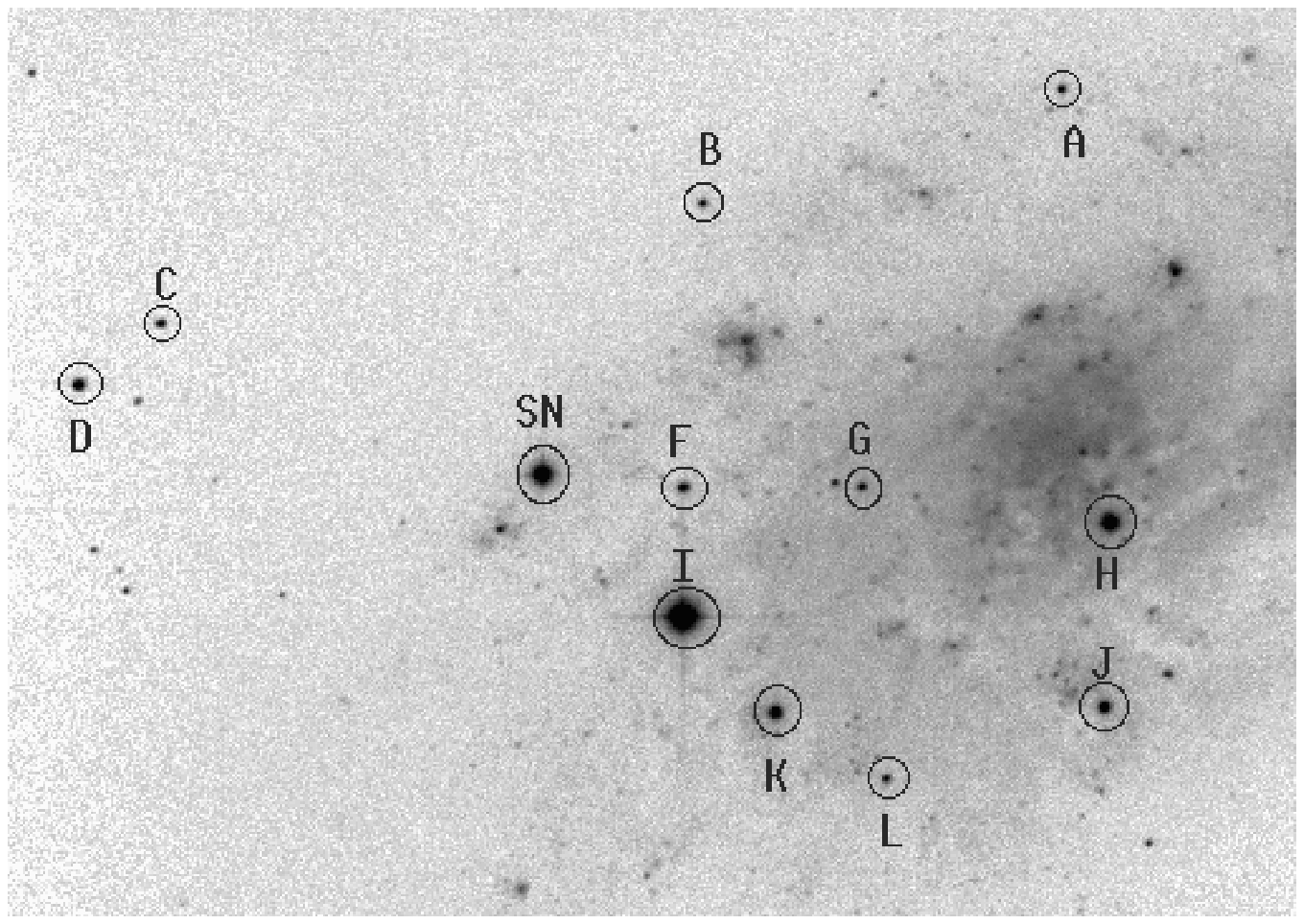,width=9cm}
\psfig{file=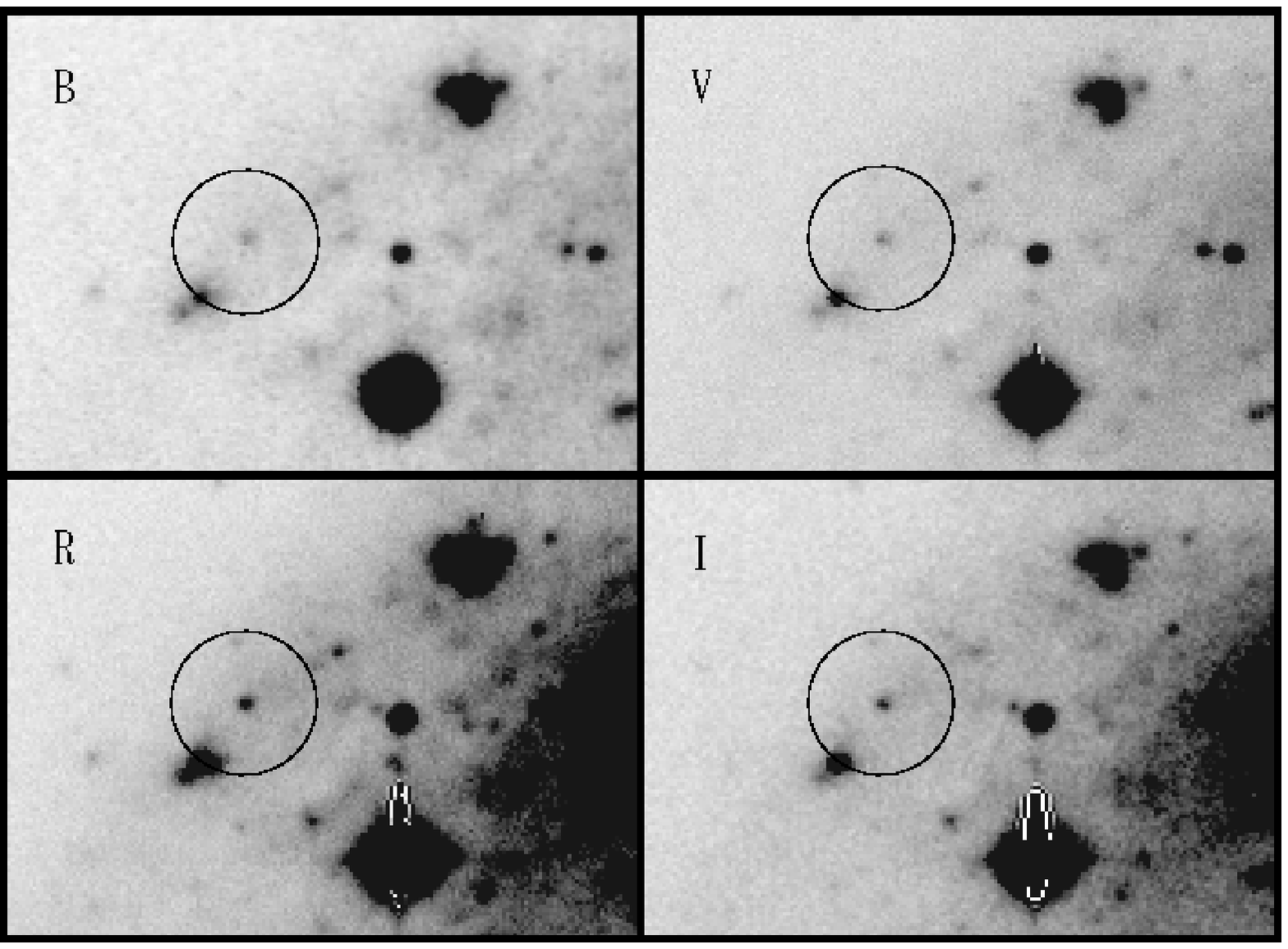,width=8.5cm}
\caption{Left: The field of SN 2004dj observed with the Princeton VersArray 1300B 
CCD on the 1m RCC telescope at Konkoly Obs.
The field of view shown is about 7x5 arcmins. North is up and East is to the left. 
The local comparison stars and the SN are labelled. The standard magnitudes 
of the comparison stars are listed in Table 2. Right: The detection of the progenitor
cluster Sandage 96 on archival frames taken with the 60/90 cm Schmidt telescope and 
Photometrics CCD at Konkoly Obs. on 11th Jan. 2002. The applied filter is indicated on the
upper left corner on each frame.}
\end{center}
\end{figure*}

Type II supernovae (SNe) emerge from shock-driven explosion of massive ($M ~>~ 8 ~ M_{\odot}$) 
stars initiated by core collapse \citep{ww,nady}. In particular, plateau Type II SNe 
(SNe IIp) result from core collapse of supergiants that have massive hydrogen-rich envelope.
The light curves of SNe IIp are characterized by a constant luminosity plateau lasting about 
80 - 120 days after explosion \citep{nady}. 
Recently, the progenitors of a few SNe IIp have been directly identified on pre-explosion images
\citep{maund1, vandyk1, li1, maund2, li2}, revealing that these are $8 - 15 ~M_{\odot}$ supergiants,
close to the low-mass theoretical limit of the core collapse process.

SN~2004dj in NGC~2403 is an outstanding SN discovered by Itagaki \citep{nakano} 
on July 31, 2004 (JD 2453218), because it is
one of the brightest and closest supernova observed ever. Based on its first optical spectrum \citep{patat1}
SN~2004dj was quickly classified as a "normal" Type IIp event. Unfortunately, it was discovered during the
plateau phase ($\sim 1$ month after explosion), thus, the maximum of the light curve could not be covered. 
Due to its proximity ($D \sim 3.1$ Mpc) \citep{freedman1}, SN~2004dj was also detected in radio 
\citep{stock,besw}, infrared \citep{suger,kotak} and X-ray bands \citep{pool}. 
$R$-band optical photometry and high-resolution $H \alpha$ spectroscopy obtained during the plateau 
phase was published by \citet{korcak}. \citet{chugai} presented $BVR$ light curve and full optical spectra 
in the nebular phase. They estimated the amount of $^{56}$Ni as $\sim 0.02 ~M_{\odot}$, and pointed out an
asymmetric, bipolar distribution of the ejected nickel from modeling the nebular $H \alpha$ profile.
The optical light curve observed in the BATC (intermediate-band) system was also studied 
by \citet{zhang}. They concluded that SN~2004dj was a typical SN IIp that ejected a 
$\sim 10~M_{\odot}$ envelope and synthesized $\sim 0.02 ~M_{\odot}$ of radioactive nickel. 

Particular interest was paid on the association of SN~2004dj 
with the compact cluster Sandage~96 which is thought to be the host of the progenitor star \citep{yama}. 
\citet{maiz} and \citet{wang} analysed the spectral energy distribution (SED) of Sandage~96 
(based on archival CCD photometry), and concluded
that S96 is indeed a compact cluster with age of $\sim 14 - 20$ Myr and total stellar mass of 
$\sim 10^4 - 10^5 ~ M_{\odot}$ (the uncertainty is due to the different reddening determination). 
From population synthesis, the initial mass of the progenitor star of SN~2004dj was estimated as 
$12 - 15 ~ M_{\odot}$, which is in the upper part of the mass range of other 
SNe IIp progenitors (see above). 

In this paper we present time-resolved optical photometry and spectroscopy 
of SN~2004dj covering its first year
after discovery. Our primary goal is to derive a reasonable distance to this SN and 
its host galaxy
NGC~2403 by using all available information. The new observational data are presented 
in Section 2 and 3. 
In Section 4 we apply a variant of the Expanding Photosphere Method (EPM) and the SNe IIp 
Standard Candle Method 
(SCM) for distance determination, and critically compare the results with other 
distance estimates of the host galaxy.
Based on the improved distance, we estimate some of the physical properties of 
SN~2004dj and S96 in Section 5. Finally,
Section 6 summarizes our results.

\section{Observations}
\subsection{Photometry}

\begin{table*}
\begin{center}
\caption{Basic data of telescopes used for photometric observations. The columns contain
the followings: code, observatory, diameter of telescope, manufacturer of CCD, size of CCD
in pixels, readout noise in electrons, pixel scale in arcseconds, field of view on the CCD in
arcminutes and typical FWHM of stellar PSFs in arcseconds. Note that the CCD at Szeged 
Observatory was used with 2x2 binning.}
\begin{tabular}{lcccccccc}
\hline
Code & Observatory & Telescope & CCD & Pixels & RON (e$^-$)& Scale (") & FOV (') & FWHM (") \\
\hline
A&Konkoly& 60/90 cm Schmidt & Photometrics & 1024x1536 & 14 & 1.00 & 17x25.5 & 3.3 \\
B&Konkoly& 100 cm RCC & Princeton VersArray & 1340x1300 & 10? & 0.30 &  6.7x6.5 & 3.0 \\
C&Szeged& 40 cm Cass & SBIG ST-9E & 512x512 & 13 & 0.70 & 6x6 & 3.8 \\
D&FLWO& 120 cm & Minicam & 4800x4800 & 7.6 & 0.30 & 24x24 & 1.2 \\
\hline
\end{tabular}
\end{center}
\end{table*}

The photometric data were collected with four telescopes, whose basic parameters are summarized in Table~1.
All observations were made through Johnson-Cousins $BVRI$ filters. 
The brightness of the SN was tied to local 
comparison stars, whose magnitudes were calibrated via Stetson's standard field of NGC 7790, when the 
sky was photometric. These local comparison stars are shown in Fig.~1.

The reduction of the CCD frames was performed in $IRAF$ following the standard procedure: bias/dark subtraction 
and flat field division, The photometry of the reduced frames was computed by two methods. First, 
the instrumental magnitudes were derived via aperture photometry. The aperture radius
was set as $2 \times$FWHM of the stellar images. The local background was determined in an annulus 
with inner radius and
width of $3 \times$FWHM and $2 \times$FWHM, respectively. Because of the galaxy contamination on the
SN and several other local calibrator stars, the background flux around each star has been determined with the
built-in {\it centroid} algorithm that is expected to give the best estimate on quickly variable background.
In order to test this result, the background level was also calculated with the {\it mode} algorithm
($3 \times$median - $2 \times$mean), after eliminating the outliers that 
deviated more than $3 \sigma$ from the mean value. No significant difference was found between the
two background levels, suggesting that the background determination around the selected objects is
robust. 

The stellar magnitudes coming from the aperture photometry 
were checked with the results of simultaneous PSF-photometry. This was done only for the 
Konkoly Schmidt telescope frames, where the larger field of view contained 
25 - 30 bright stars, thus, reliable PSF could be determined.
This comparison is shown in Fig.~2, where the aperture minus PSF magnitudes are plotted
against the $V$ magnitude. 
It can be seen that most of the magnitudes of the two photometric methods agree well 
within $\pm 0.05$ mag.
A few outliers correspond to the faintest stars in the $B$ band. It is concluded that the results
of the aperture photometry with the parameters given above are in accord with the PSF-magnitudes,
therefore, the systematic error caused by the photometric method is negligible with respect to
the photon/detector noise and errors of the standard transformation.

\begin{figure}
\begin{center}
\leavevmode
\psfig{file=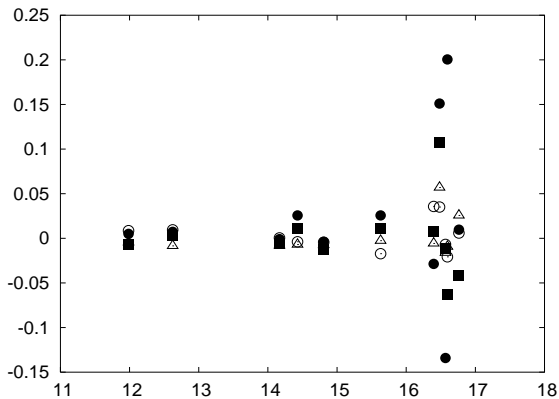,width=8cm}
\caption{Difference of the magnitudes obtained from aperture- and PSF-photometry against
the $V$ magnitude. The symbols denote the filters as follows: filled circles - B, filled squares - V, 
open circles - R, open triangles - I filter.}
\end{center}
\end{figure}

The host galaxy NGC 2403 was also observed in $BVRI$ bands with the Schmidt telescope 
at Konkoly Observatory 
during the course of our nearby galaxy survey programme\footnote{http://astro.u-szeged.hu/\~{}sn} 
\citep{sry} on 23 December 2001 and 11 January 2002. 
The progenitor cluster Sandage~96 is clearly present on these frames (Fig.~1 right panel). 
Using the same set of comparison stars, the magnitudes of S96 were computed in the same way as 
described above. The results are in the last row of Table~2. Our magnitudes are in very good 
agreement with those presented by \citet{maiz}. Because most of the light of the progenitor cluster
is still present after the SN explosion, the measured magnitudes of SN~2004dj have been corrected
for the light of the progenitor in the following way: the magnitudes of the progenitor as well as
those for the SN plus progenitor have been converted into fluxes, then the progenitor fluxes
have been subtracted from the SN+progenitor fluxes. The resulting fluxes, expected to be
due to only the SN, were transformed back into magnitudes. 

\begin{table}
\begin{center}
\caption{Standard magnitudes of local comparison stars. Errors are given in parentheses. 
See Fig.~1 for object identification. The last row lists our measurements of the progenitor
cluster Sandage~96.}
\begin{tabular}{lcccc}
\hline
Star & $V$ & $B-V$ & $V-R$ & $V-I$ \\
\hline
A & 16.60(0.05) & 0.71(0.10) & 0.42(0.06) & 0.77(0.05) \\
B & 16.80(0.07) & 1.57(0.09) & 0.97(0.05) & 1.84(0.05) \\
C & 16.34(0.06) & 0.89(0.07) & 0.57(0.04) & 0.97(0.03) \\
D & 14.81(0.04) & 0.76(0.02) & 0.45(0.02) & 0.78(0.02) \\
F & 15.61(0.05) & 0.95(0.05) & 0.58(0.05) & 0.97(0.03) \\
G & 16.42(0.05) & 0.90(0.06) & 0.53(0.04) & 0.93(0.04) \\
H & 12.61(0.03) & 0.98(0.04) & 0.56(0.03) & 1.00(0.02) \\
I & 10.06(0.03) & 0.56(0.02) & 0.38(0.02) & 0.64(0.02) \\
J & 14.46(0.04) & 0.70(0.04) & 0.43(0.03) & 0.78(0.02) \\
K & 14.20(0.04) & 0.74(0.04) & 0.45(0.03) & 0.78(0.02) \\
L & 16.66(0.05) & 0.88(0.09) & 0.54(0.06) & 0.92(0.06) \\
S96 & 17.85(0.05) & 0.40(0.07) & 0.33(0.06) & 0.79(0.05) \\
\hline
\end{tabular}
\end{center}
\end{table}

\begin{figure}
\begin{center}
\leavevmode
\psfig{file=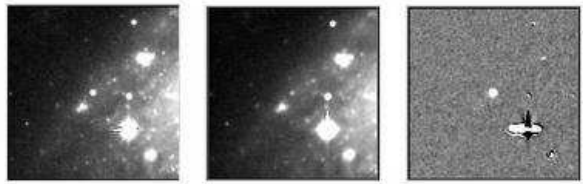,width=8cm}
\caption{Results of the image subtraction method illustrated with the $R$ frames obtained on 
May 5, 2005. Left panel: the broadened object frame after convolving with the kernel produced by
$psfmatch$. Middle panel: the template frame scaled to the light level of the object frame. Right
panel: the difference between the two images. The galaxy and the unsaturated stars have been 
removed.}
\end{center}
\end{figure}

\begin{figure}
\begin{center}
\leavevmode
\psfig{file=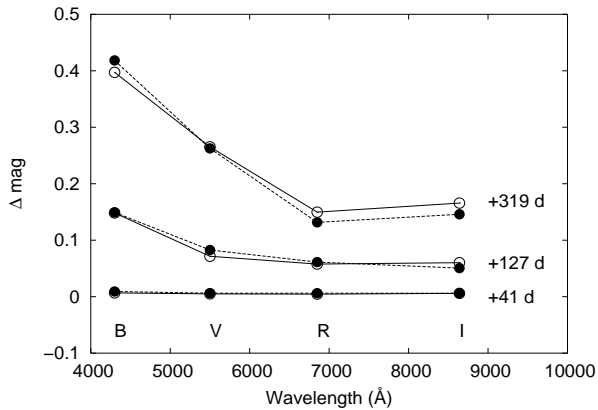,width=8cm}
\caption{Magnitude corrections due to the progenitor cluster computed with $i)$ direct aperture photometry of the
SN and template frames (filled circles) and $ii)$ image subtraction method (open crcles) as a function of
wavelength. The phase of the observation is labelled on the right. The magnitude corrections from the
two methods agree within $0.02$ mag, even at the last epoch, when the SN was the faintest on our frames. }
\end{center}
\end{figure}

In order to further test the reliability of the background subtraction and the correction 
for the progenitor, the Konkoly Schmidt frames were also analysed using the image subtraction method.
This was also done with built-in $IRAF$ tasks. Three epochs have been selected: Aug.8, 2004 ($+39$ days after
explosion), Nov.02, 2004 ($+125$ days) and May 13, 2005 ($+317$ days), when the SN was in the middle of
the plateau phase, during the transition phase and in the nebular phase, respectively. 
The template frames containing the progenitor have been subtracted from the $BVRI$ frames obtained on these epochs 
with the Schmidt telescope in the following way. First, the images have been registered with the $geomap$,
$geotran$ and $imalign$ tasks. Second, the stellar PSFs on the object and template frames have been matched
using the $psfmatch$ task. Because the PSFs of the template frames were broader than those of the object frames,
the latter images have been broadened with the kernel produced by $psfmatch$. Third, the intensities of
the comparison stars have been scaled to the same level on both frames with the task $linmatch$. Fourth,
the scaled template frames have been removed from the broadened object frames to produce the subtracted
images. The result of this procedure is illustrated in Fig.3. It is seen that the galaxy and the unsaturated
stars have been nicely removed by the image subtraction, leaving a well-defined object (the SN) on a
flat background. Then, the aperture photometry have been computed again, now measuring the comparison
stars on the broadened object frames, and the SN on the subtracted frames. The SN magnitudes obtained
this way were compared with the results of the first method, i.e. that were computed directly 
on the object frames and corrected for the progenitor magnitudes (see above). Fig.4 shows the 
difference between the uncorrected and progenitor-corrected SN magnitudes in the case of direct aperture
photometry (filled circles), and the difference between the SN magnitudes measured before and after
the template subtraction (open circles), as a function of wavelength, for the three epochs considered.
It is visible that the corrections coming from the direct 
photometry do agree within $0.02$ mag with those from the image subtraction method, 
even at the last epoch when the SN was the faintest on our frames. 
This also means that the galaxy contamination could also be effectively removed with the 
selected aperture/annulus combination in the first method.
Note that this good agreement is partly due to the fact that
the object and template frames have been made with the same instrument. The image subtraction method
is expected to perform less well for those SN frames that have been made with other telescopes.
Moreover, the digital image subtraction always increases the noise on the difference image, resulting
in slightly higher uncertainties. Therefore, it is concluded that the light of the
progenitor cluster can be reliably removed from the SN magnitudes obtained with direct aperture
photometry of the object frames (containing the galaxy and the SN) using the progenitor magnitudes
listed in Table~2. The image subtraction method (that removes the host galaxy and the progenitor 
pixel-by-pixel) gave essentially the same result for the selected frames. In the followings we use
the magnitudes of SN~2004dj computed with the first method and corrected for the progenitor
magnitudes in Table~2.  
    
Finally, all photometric data were transformed
into the standard Johnson-Cousins system using colour terms determined for each 
telescope/CCD on photometric nights. 
The final calibrated and progenitor-corrected magnitudes of SN~2004dj are listed in 
Table~3. The errors (given in parentheses) reflect the observational noise and the 
uncertainties of the magnitudes of local comparison stars (see Table~2).

\begin{figure}
\begin{center}
\leavevmode
\psfig{file=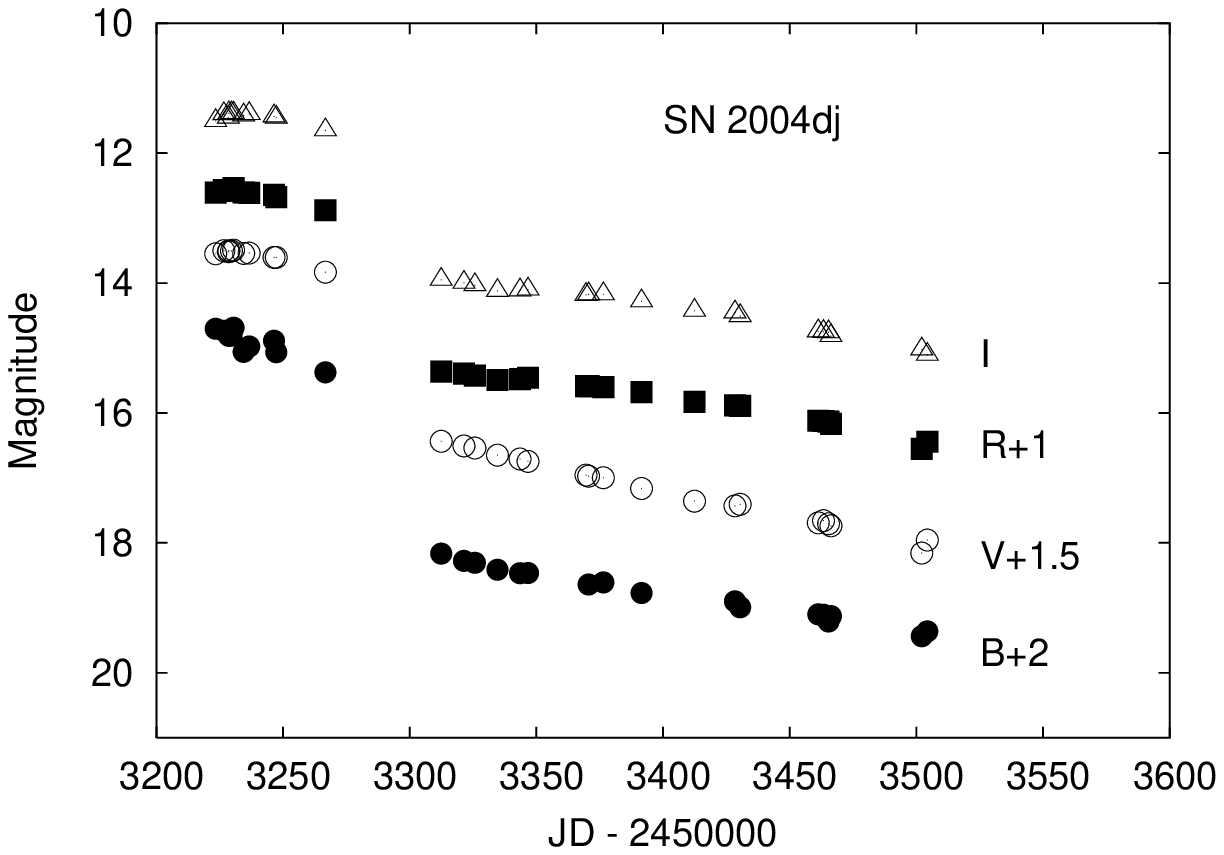,width=8.5cm}
\psfig{file=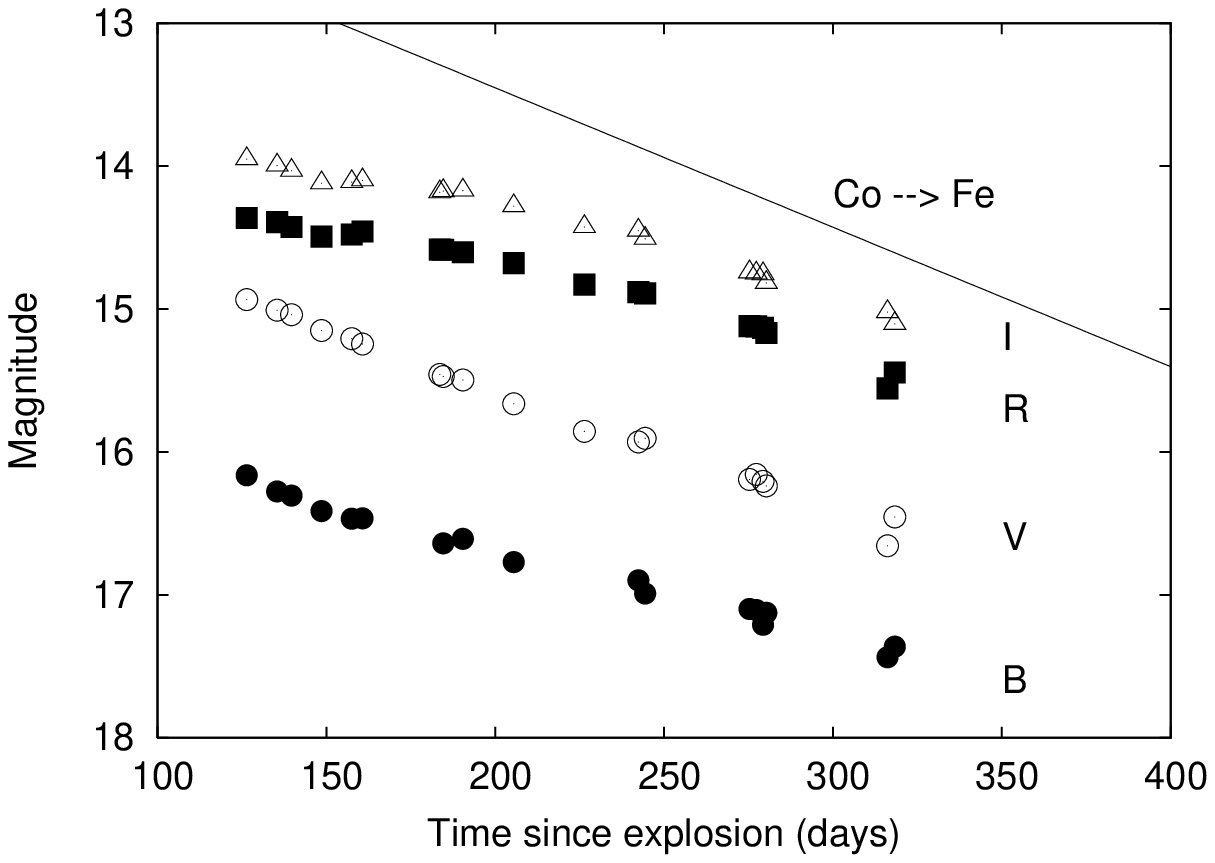,width=8.5cm}
\caption{Top panel: Light curve of SN 2004dj in $BVRI$ bands. The curves have
been shifted vertically for better visibility with the amount indicated in the
legend on the right.
Bottom panel: The light curve during
the early tail phase. The line indicates the expected slope of the Co-decay.}
\end{center}
\end{figure}

\begin{table*}
\begin{center}
\caption{Observed magnitudes of SN 2004dj. The columns contain the followings: 
date and JD-2450000 of the observations, days relative to the estimated
moment of explosion (JD 2453187, see Sect.4), $BVRI$ magnitudes corrected for
the magnitudes of S96 (errors are in parentheses)
and codes of the applied telescopes. See Table 1 for the telescope codes.}
\begin{tabular}{lccccccc}
\hline
Date & JD-2450000 & $t - t_{expl}$ & $B$ & $V$ & $R$ & $I$ & Tel.\\
\hline
2004-08-05 & 3223.4 & 36& 12.70(0.05)& 12.05(0.04)& 11.61(0.03)& 11.50(0.07)& C\\
2004-08-08 & 3226.6 & 39& 12.73(0.06)& 12.00(0.03)& 11.57(0.03)& 11.40(0.03)& A\\
2004-08-10 & 3228.4 & 41& --	     & 12.02(0.03)& 11.57(0.03)& 11.45(0.05)& C\\
2004-08-10 & 3228.6 & 41& 12.81(0.05)& 12.00(0.03)& 11.58(0.03)& 11.39(0.03)& A\\
2004-08-11 & 3229.6 & 42& 12.81(0.05)& 12.01(0.03)& 11.58(0.03)& 11.39(0.03)& A\\
2004-08-12 & 3230.4 & 43& 12.69(0.05)& 11.99(0.03)& 11.54(0.05)& 11.39(0.03)& C\\
2004-08-16 & 3234.4 & 47& 13.06(0.07)& 12.04(0.05)& 11.61(0.03)& 11.42(0.03)& C\\
2004-08-18 & 3236.6 & 49& 12.98(0.06)& 12.04(0.03)& 11.61(0.03)& 11.40(0.03)& B\\
2004-08-28 & 3246.4 & 59& 12.89(0.12)& 12.10(0.03)& 11.64(0.03)& 11.43(0.03)& C\\
2004-08-29 & 3247.3 & 60& 13.07(0.06)& 12.11(0.03)& 11.68(0.03)& 11.45(0.03)& C\\
2004-09-17 & 3266.7 & 79& 13.38(0.05)& 12.33(0.03)& 11.88(0.03)& 11.65(0.03)& A\\
2004-11-02 & 3312.4 &125& 16.16(0.05)& 14.93(0.04)& 14.36(0.03)& 13.95(0.03)& A\\
2004-11-11 & 3321.4 &134& 16.28(0.05)& 15.01(0.04)& 14.39(0.03)& 13.99(0.03)& A\\
2004-11-15 & 3325.7 &138& 16.31(0.05)& 15.04(0.04)& 14.43(0.03)& 14.03(0.03)& A\\
2004-11-24 & 3334.6 &147& 16.41(0.17)& 15.15(0.08)& 14.49(0.08)& 14.12(0.06)& B\\
2004-12-04 & 3343.5 &156& 16.47(0.05)& 15.21(0.04)& 14.48(0.04)& 14.11(0.03)& B\\
2004-12-07 & 3346.7 &159& 16.47(0.05)& 15.24(0.04)& 14.46(0.04)& 14.10(0.03)& B\\
2004-12-29 & 3369.6 &182& --	     & 15.46(0.06)& 14.58(0.04)& 14.18(0.03)& A\\
2004-12-30 & 3370.6 &183& 16.64(0.05)& 15.47(0.04)& 14.58(0.05)& 14.17(0.03)& A\\
2005-01-05 & 3376.4 &189& 16.61(0.05)& 15.50(0.04)& 14.60(0.05)& 14.17(0.03)& A\\
2005-01-21 & 3391.5 &204& 16.77(0.05)& 15.66(0.04)& 14.68(0.04)& 14.28(0.03)& A\\
2005-02-10 & 3412.4 &225& --	     & 15.86(0.06)& 14.83(0.04)& 14.43(0.05)& A\\
2005-02-26 & 3428.4 &241& 16.90(0.06)& 15.93(0.04)& 14.88(0.05)& 14.45(0.05)& A\\
2005-02-28 & 3430.4 &243& 16.99(0.06)& 15.90(0.04)& 14.89(0.04)& 14.51(0.03)& A\\
2005-03-31 & 3461.3 &274& 17.10(0.07)& 16.19(0.05)& 15.12(0.04)& 14.74(0.04)& A\\
2005-04-02 & 3463.3 &276& 17.11(0.07)& 16.16(0.05)& 15.12(0.04)& 14.75(0.04)& A\\
2005-04-04 & 3465.3 &278& 17.21(0.07)& 16.21(0.05)& 15.13(0.04)& 14.76(0.04)& A\\
2005-04-05 & 3466.3 &279& 17.13(0.08)& 16.24(0.05)& 15.17(0.04)& 14.82(0.04)& A\\
2005-05-12 & 3502.1 &315& 17.44(0.10)& 16.66(0.05)& 15.55(0.04)& 15.02(0.04)& D\\
2005-05-13 & 3504.3 &317& 17.36(0.10)& 16.46(0.05)& 15.44(0.04)& 15.10(0.04)& A\\
\hline 
\end{tabular}
\end{center}
\end{table*}

The light curves of SN~2004dj are plotted in Fig.~5. This is a typical Type IIp light curve.
During the plateau phase the light level is almost constant in $V$, $R$, $I$ bands. There
is a slow decrease in $B$, which would be even stronger in the $U$ band. It is generally
accepted that this plateau phase is due to a hydrogen recombination front that moves inward
the expanding H-rich ejecta. 
Then, following a rapid transition, the SN enters the nebular
phase (often referred as the tail phase), where the light variation 
reflects the radioactive decay of $^{56}$Co into $^{56}$Fe.
The bottom panel of Fig.~5 shows only the radioactive tail with the expected slope of the
Co-decay. It is visible that in the early tail phase the slope of the light curve is less 
than that of the Co-decay in $R$ and $I$. The decline rate in these bands is 
0.6 mag/(100 days) compared with the 0.98 mag/(100 days) rate of the Co-decay. 
On the other hand, the $B$ and $V$ light curve declines with 0.8 mag/(100 days) which
is close to the expected rate. Similar phenomenon was found by \citet{zhang} from
their intermediate-band light curves: in the $\lambda \lambda 6600 - 8500$ \AA~ 
wavelength regime the decline rates are much less than the Co-rate. Since the
$R$ and $I$ bands are dominated by the $H \alpha$ and Ca-triplet emission lines,
which can be particularly strong at the beginning of the tail phase, it is
probable that the reduced decline rate in these bands reflects the contribution
of these lines to the thermal radiation powered by the trapping of gamma-rays 
and positrons from the Co-decay. 

\citet{chugai} also presented a tail light curve of SN~2004dj 
obtained through Johnson-Cousins $B$, $V$ and $R$ filters. 
Comparing their published photometry with ours presented here, the light curve 
slopes in different bands are consistent. However, there is a general, systematic offset
of 0.3 - 0.4 mag in all bands in the sense that the Chugai et al. magnitudes are
systematically brighter. Both sets of light curves have been transformed to the standard
system, although there is no star in common in the two groups of local calibrators.
Since \citet{chugai} do not mention whether they had corrected their data for the
progenitor light, it is probable that this difference is partly due to the presence
of the progenitor light in their magnitudes. But it is interesting that if we
do not correct our data for the progenitor, the offset is still present, although
with a reduced amount (Chugai et al.'s data are still brighter by $\sim 0.2$ mag). 
At present, it is not clear what is the cause of this systematic offset, but note 
that the standard transformation of SN magnitudes obtained with different 
telescope/CCD/filter combinations are always uncertain at the level of $\sim 0.1$ mag
\citep{suntzeff} due to the nonstellar SED of the SN, especially during the nebular
phase. This effect may easily be responsible for most of the remaining offset.

\subsubsection{Comparison with other SNe Type IIp}

\begin{figure}
\begin{center}
\psfig{file=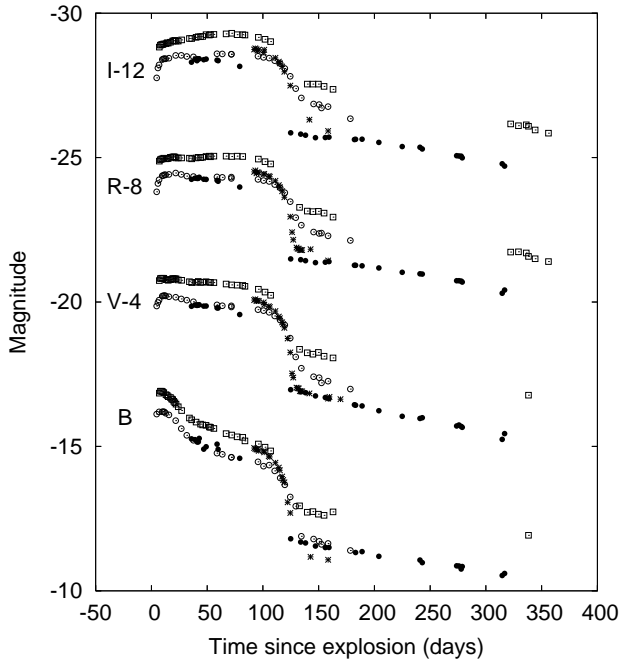,width=8.5cm}
\caption{Comparison of the light curves of SN 2004dj (filled circles), SN 1999gi (open
circles), SN 1999em (open squares) and SN 2003gd (asterisks). 
The magnitudes have been corrected for reddening and distance listed in Table 4. 
The curves have been shifted vertically for better visibility (the amount of the applied
magnitude shift is indicated on the left-hand side of each curve}.
\end{center}
\end{figure}

It is interesting to compare the light curve of SN~2004dj with those of other Type IIp SNe, 
although it is known that SNe IIp light curves are quite heterogenious. Recently \citet{hp} 
and \citet{hamuy1}
showed that there is a correlation between the absolute magnitudes in the middle 
of the plateau and the photospheric radial velocity at the same epoch, which partly explains
the observed differences in the light curves of different SNe IIp. 

\citet{chugai} compared SN~2004dj with SN~1999gi, another well-observed Type IIp SN. 
Based on the $V$ light curve (supplemented by amateur observations) and an estimated 
explosion date JD 2453170, they concluded that the two light curves are very similar. 
\citet{zhang} did the same but using the data of SN~1999em for comparison. Based on the
assumption that the two light curves should be similar, they estimated
the moment of explosion as JD $2453167 \pm 21$, which was nearly the same as \citet{chugai}
determined. 
  
We compared our new light curve (Fig.5 and Table~3) with the published magnitudes of
SN~1999em \citep{leonard1} , SN~1999gi \citep{leonard2} and SN~2003gd \citep{hendry} 
To do this, all data were
corrected for reddening and distance modulus, collected from references listed
in Table~4. For SN~1999em we have used the distance derived recently by \citet{dh2},
which is in good agreement with the Cepheid-based distance of its host galaxy.
For SN~2004dj we have applied our preferred values of $E(B-V) = 0.07$ mag, 
$T_{expl} = 2453187$  and $D = 3.47$ Mpc (see Section 4). 
The absolute magnitudes for each SN are plotted together in Fig.6.

\begin{table}
\begin{center}
\caption{Light curve parameters of Type IIp SNe used for comparison.}
\begin{tabular}{lcccr}
\hline
SN & E(B-V) & $t_{expl}$ & $D$ & Ref. \\
   & (mag) & (JD-2450000) & (Mpc) & \\
\hline
1999em & 0.10 & 1475& 11.5 & 1,3 \\
1999gi & 0.22 & 1518& 10.8 & 2 \\
2003gd & 0.14 & 2717& 9.3 & 4 \\
2004dj & 0.07 & 3187& 3.47 & 5 \\
\hline
\end{tabular}
\end{center}
References - (1) \citet{leonard1}; (2) \citet{leonard2}; (3) \citet{dh2}; 
(4) \citet{hendry}; (5) present paper \\
\end{table}
 
From Fig.6 it is visible that the absolute light level of SN~2004dj is similar
to that of SN~1999gi during the plateau phase, but different from those of SNe~1999em and
2003gd. This difference may be due to different nickel masses for these SNe. 
However, as it will be discussed in Section 2.2, the spectra and the expansion 
velocities are quite similar for SN~2004dj and SN~1999em in the middle of the
plateau phase, that would suggest similar plateau magnitudes \citep{hp}.
It seems that the plateau phase of SN~2004dj ended 
slightly early relative to 99em and 99gi. Comparing our data with the $R$ light curve of
\citet{korcak}, we estimate the starting date of the transition into the nebular phase 
as $t ~=~ +80 ~\pm 20$ days (the uncertainty is due to both the poor sampling of
the light curves around this epoch and the uncertainty of the explosion date). 
The inflection time \citep{elmha1} for SN~2004dj is estimated as $t_i = +110 \pm 20$ 
days, which is slightly earlier than that for SN~1999em ($+120 \pm 4$ days) 
and SN~1999gi ($+125 \pm 3$ days), but this is not a significant difference 
because of the relatively high uncertainty of the explosion date of SN~2004dj.

In the nebular phase, SN~2004dj has the faintest light level among these four SNe,
except in $B$ where SN~2003gd is slightly fainter. The faintness of SN~2004dj is 
increasing toward longer wavelengths. In this phase SN~2004dj is more
similar to SN~2003gd \citep{vandyk1, hendry} than the other two SNe.
\citet{elmha1} found that the tail light level of SNe IIp correlate well 
with the plateau magnitude in $V$. From Fig.~6 it is suspected that this 
correlation is actually weak in $V$, and probably not valid at all in $R$ and $I$.

Since SNe IIp tail light curves are powered by radiactive decay
of $^{56}$Co - $^{56}$Fe, different brightness probably reflects the
difference in the amount of the synthesized iron-peak elements, particularly $^{56}$Ni
\citep{nady, elmha1}. From hydrodynamic models 
\citet{nady} pointed out that the middle-plateau absolute magnitude $M_V$ is proportional to 
the plateau duration $\Delta t_p$, the expansion velocity 
in the middle of the plateau and the explosion energy (see Eq.1 of \citet{nady}). 
Thus, from the shape of the light curve alone,
we suspect that SN~2004dj had similar explosion energy and nickel mass than those of SN~1999gi,
but less than those of SN~1999em. We will examine this question more 
quantitatively in Section~5.

\subsubsection{Reddening}

\begin{figure}
\begin{center}
\psfig{file=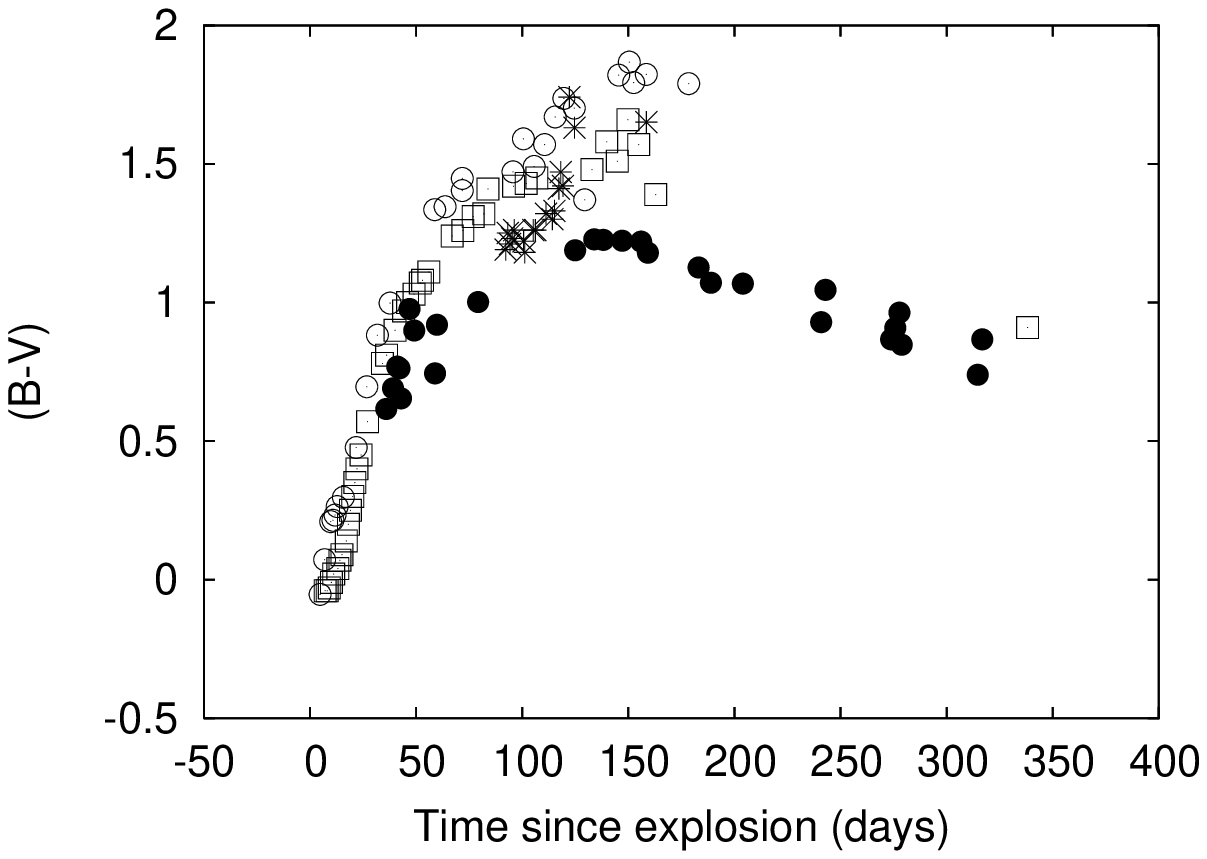,width=6cm}
\psfig{file=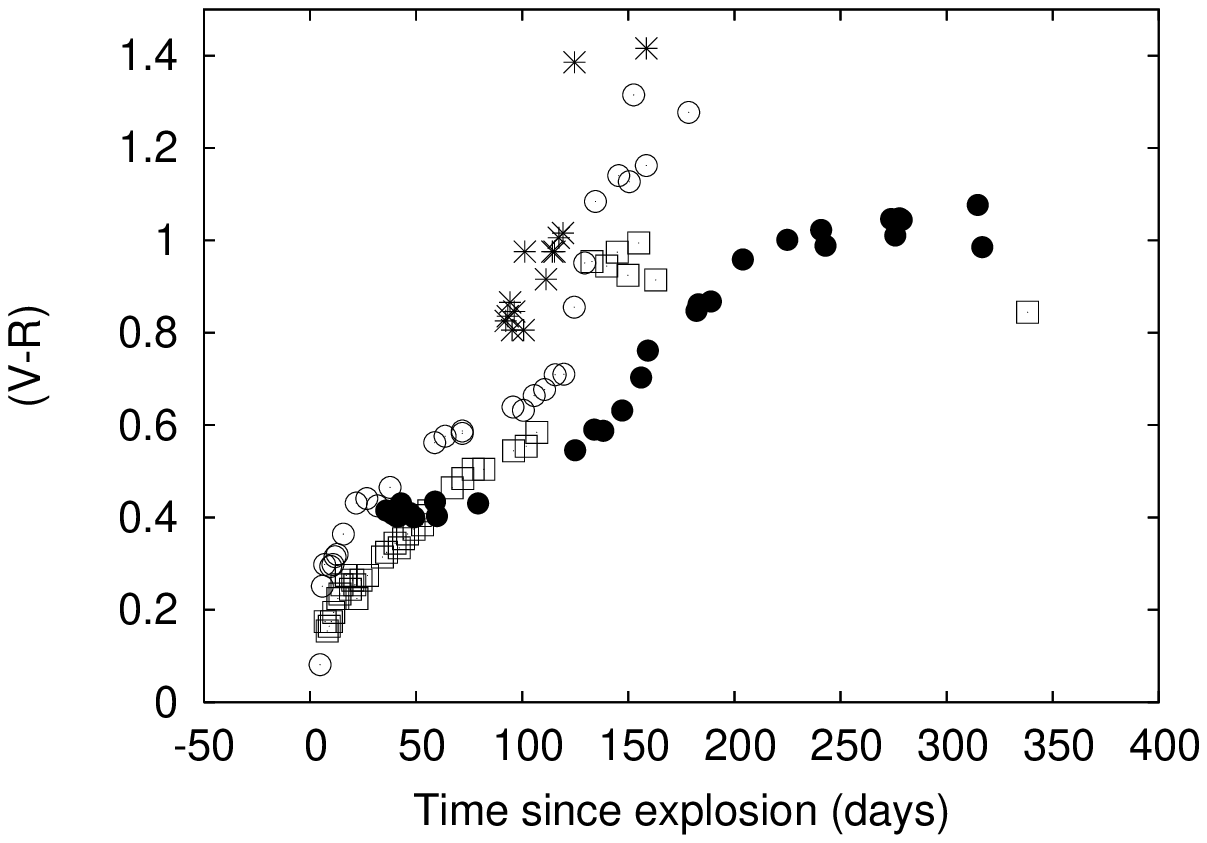,width=6cm}
\psfig{file=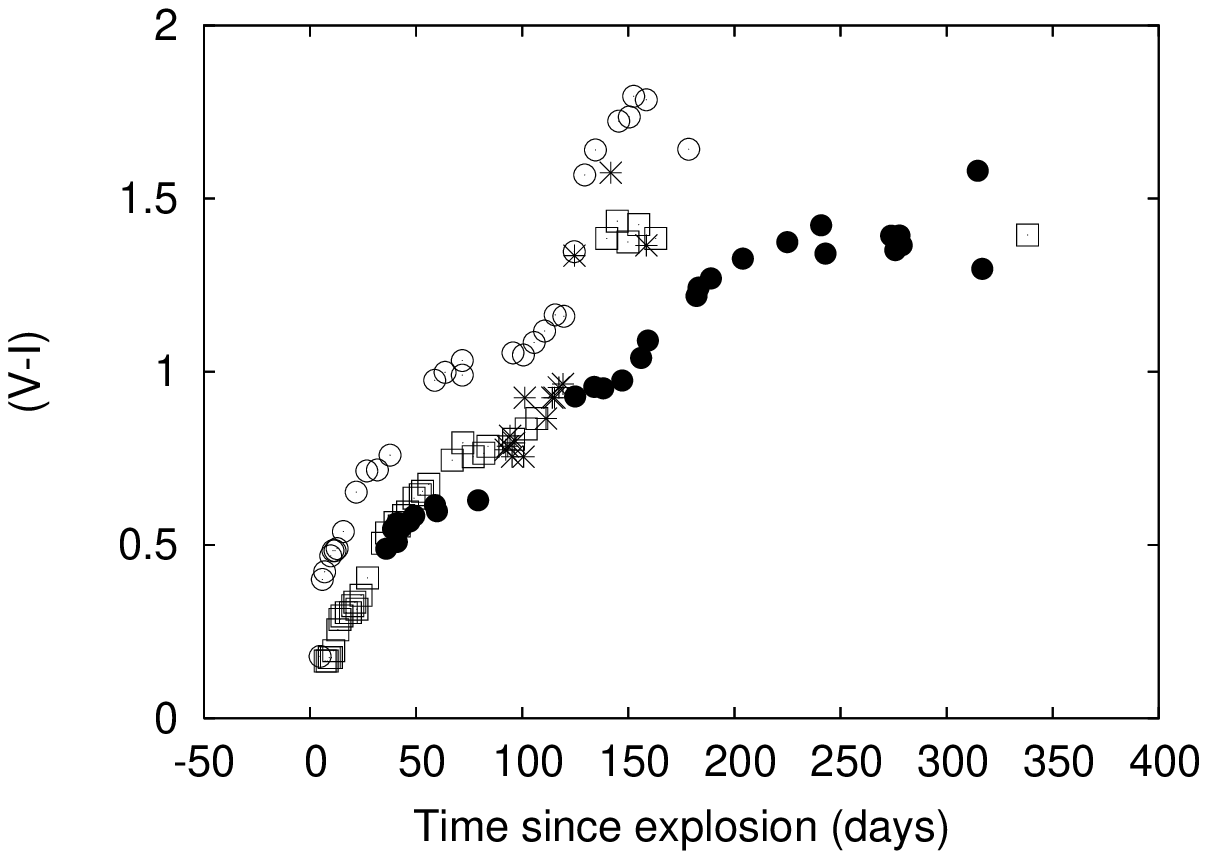,width=6cm}
\caption{The colour curves of SNe~2004dj, 1999em, 1999gi and 2003gd corrected for
galactic reddening (see text). The symbols are the same as in Fig.~6.}
\end{center}
\end{figure}

Reddening is the key parameter when photometry is used to derive physical parameters. 
From the literature, the reddening of SN~2004dj is controversial. \citet{patat1}
reported the identification of interstellar Na D (from an unresolved absorption trough)
in the very first spectrum of SN~2004dj, from which they deduced $E(B-V)_{total} = 0.18$ mag.
1 day later \citet{gk} stated that with high-resolution echelle spectroscopy they
resolved the Na D lines in the host galaxy NGC~2403. They revealed the contribution of
the host galaxy to the reddening of SN~2004dj as $E(B-V)_{host} = 0.026$ mag. Since the galactic
component is only $E(B-V)_{gal} = 0.04$ mag \citep{sfd}, this indicates significantly less
total reddening ($E(B-V)_{total} = 0.066$) than \citet{patat1} estimated. 

Reddening has also been determined indirectly from population synthesis models of the 
progenitor cluster S96. Fitting cluster SEDS to archival optical and near-IR 
($UBVRIJHK$) photometric data of S96, \citet{maiz} derived 
$E(B-V)_{total} \approx 0.13 \pm 0.03$ mag. This value is between the two direct spectral
measurement mentioned above, being definitely in better agreement with the result of 
\citet{patat1}. On the other hand, \citet{wang} have applied essentially the
same technique using their own O-IR photometric data (through intermediate-band
filters, providing better wavelength coverage in the optical) on S96. They have
obtained $E(B-V)_{total} = 0.35 \pm 0.05$ mag, arguing that this value provides a better fit
of the calculated SEDs. \citet{zhang} also derived $E(B-V)_{total} = 0.33 \pm 0.11$ mag
from the reanalysis of the EW of the interstellar Na D trough reported by \citet{patat1}.
It is clear that the present estimates of the extragalactic reddening
toward SN~2004dj range from an almost negligible value \citep{gk} to 
significant in-host extinction \citep{wang,zhang}. 

Comparing our observed colour curves of SN~2004dj with those of SNe~1999em, 1999gi and 2003gd 
some constraints can be determined on the upper or lower limits of reddening \citep{leonard2}. 
Fig.~7 shows the $B-V$, $V-R$ and $V-I$ curves of these four SNe IIp corrected for the galactic
reddening (the following values were adopted from NED: 
$E(B-V)_{gal} = 0.04, ~0.04, ~0.017$ and $0.069$ mag for SNe 2004dj, 1999em, 1999gi and 2003gd, respectively).
SN~2004dj has the bluest colours among these SNe, especially in 
the transition and the nebular phase. Since SNe 1999em, 1999gi and 2003gd suffered only 
moderate reddening in their hosts (see Table~4), the bluer colour of SN~2004dj means 
that $E(B-V)_{total} > 0.1$ mag is less probable. Thus, the observed colour 
curves favour lower reddening, such as that determined by \citet{gk}. This may also be preferred, 
because this is the only one that is based on {\it direct} measurement of the resolved 
interstellar Na D lines within the host galaxy. Note that \citet{chugai} also selected
$E(B-V)_{total} = 0.062$ mag for SN~2004dj based on their observed colour indices, 
but they attributed it entirely to galactic extinction. 

In the followings we adopt $E(B-V)_{total} = 0.07 \pm 0.1$ mag as the most probable total 
(galactic plus host) reddening of SN~2004dj. Its uncertainty is estimated from the
scattering of the spectroscopic measurements mentioned above. 

\subsubsection{The bolometric light curve}

\begin{figure}
\begin{center}
\psfig{file=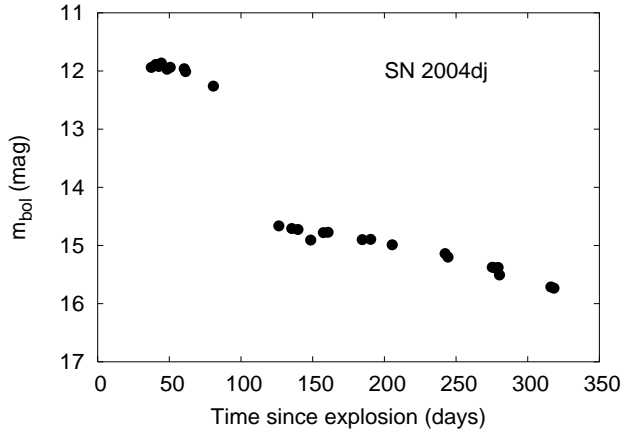,width=8.5cm}
\caption{$UVOIR$ bolometric magnitudes of SN 2004dj, corrected for reddening 
($E(B-V) = 0.07$ mag). The bolometric fluxes (Table 5) have been transformed 
into magnitudes via Eq.1.}
\end{center}
\end{figure}

\begin{table}
\begin{center}
\caption{$UVOIR$ bolometric fluxes and magnitudes of SN2004dj.}
\begin{tabular}{cccc}
\hline
JD-2450000 & $t - t_{expl}$ & $f_{bol}$ & $m_{bol}$ \\
 & (days) & ($10^{-10}$ erg/s/cm$^2$) & (mag) \\
\hline
3223.4 & 36& 4.14 &11.94 \\
3226.6 & 39& 4.35 &11.89 \\
3228.6 & 41& 4.30 &11.90 \\
3229.6 & 42& 4.30 &11.90 \\
3230.4 & 43& 4.44 &11.87 \\
3234.4 & 47& 4.02 &11.97 \\
3236.6 & 49& 4.12 &11.95 \\
3246.4 & 59& 4.06 &11.97 \\
3247.3 & 60& 3.88 &12.02 \\
3266.7 & 79& 3.16 &12.24 \\
3312.4 &125& 0.33 &14.68 \\
3321.4 &134& 0.32 &14.74 \\
3325.7 &138& 0.31 &14.77 \\
3334.6 &147& 0.28 &14.86 \\
3343.5 &156& 0.28 &14.86 \\
3346.7 &159& 0.28 &14.86 \\
3370.6 &183& 0.25 &14.97 \\
3376.4 &189& 0.25 &14.97 \\
3391.5 &204& 0.23 &15.08 \\
3428.4 &241& 0.19 &15.26 \\
3430.4 &243& 0.19 &15.31 \\
3461.3 &274& 0.15 &15.53 \\
3463.3 &276& 0.15 &15.53 \\
3465.3 &278& 0.15 &15.56 \\
3466.3 &279& 0.14 &15.59 \\
3502.1 &315& 0.11 &15.86 \\
3504.3 &317& 0.11 &15.86 \\
\hline
\end{tabular}
\end{center}
\end{table}

The $UVOIR$ bolometric light curve has been calculated from the observed $BVRI$ magnitudes
plus extrapolating an assumed (over-)simplified SED to the wavelength bands not covered by
the observations. First, the observed magnitudes were corrected for extinction using 
$E(B-V) = 0.07$ mag (see previous section) and the standard galactic reddening law 
\citep{sfd}. $K-$corrections were neglected, because the redshift of the host galaxy 
is very small, $z = 0.00044$ (NED). 
The dereddened magnitudes were then transformed to fluxes using the calibration
given by \citet{hamuy}. The fluxes have been integrated numerically using the effective 
wavelengths and FWHM values of the $BVRI$ filters by a simple midpoint rule. The fluxes 
in the missing UV and IR bands were extraplolated linearly from the $B$ and $I$ fluxes 
assuming zero flux at 3400 \AA~ and 23000 \AA. 

This approximation has been tested by
integrating the known $BVRI$ magnitudes of Vega and the Sun \citep{hamuy} and comparing
the resulting quasi-bolometric fluxes with the observed ones. The relative error of the
bolometric flux calculated in this way turned out to be 9.6 \% and 0.6 \% (0.11 and 0.001 mag) 
for the Sun and Vega, respectively. Because during the plateau phase the SED of the SN is
similar to stellar photospheres, and the effective temperature in most cases is 
between the temperature of the Sun ad Vega ($5800 ~<~ T_{SN} ~<~ 10000$ K), 
we believe that the uncertainty of our bolometric fluxes due to the approximation applied
does not exceed significantly the level of 10 percent at the early phases. 

During the nebular
phase, the SED of the SN becomes nonstellar, therefore the comparison with stellar flux
distributions may be misleading. Fortunately, SN~2004dj was detected with the IRAC and MIPS 
instruments onboard the Spitzer Space Telescope on four epochs at the beginning of the nebular 
phase. \citet{kotak} presented the photometry made in 5 channels in the mid-IR regime (from
$3.6 \mu$ to $24 \mu$). These infrared fluxes were used to test the quality of our 
quasi-bolometric fluxes in the nebular phase. Assuming negligible reddening in the mid-IR
regime, the SED of SN~2004dj was constructed using the IR fluxes obtained on Nov.1, 2004 
\citep{kotak} and our optical fluxes observed on the same epoch (Table~3). The missing
near-IR fluxes were estimated by interpolation between the optical and mid-IR fluxes,
assuming power-law wavelength dependence. This SED was
then integrated along wavelength using the same approximation in the UV boundary as before,
but extending the IR boundary up to 8 microns. The result was $0.317 \cdot 10^{-10}$ erg/s/cm$^2$,
comparing with $0.33 \cdot 10^{-11}$ erg/s/cm$^2$ estimated from the optical data. The relative
difference in only 4 percent, which is nearly the same as it was in the photospheric phase.
Thus, it is concluded that the approximative, quasi-bolometric fluxes, estimated from $BVRI$
fluxes have about 10 percent uncertainty, both in the plateau and the early nebular phase.

Finally, the bolometric fluxes have been transformed into magnitudes using
\begin{equation}
m_{bol} ~=~ -2.5 \log_{10} \left [ f_{bol} \right ] ~-~ 11.512
\end{equation}
where the zero-point was determined from the solar luminosity and absolute bolometric
magnitude ($M_{bol \odot} = +4.72$ mag and $L_\odot = 3.847 \cdot 10^{33}$ erg s$^{-1}$ 
was used). The bolometric magnitudes are plotted against time in Fig.~8.
The slope of the light curve is clearly increasing in the radioactive tail phase
(as it was also presented in Fig.~5). At $t \approx +120$ days after explosion the slope 
is about $-0.45$ mag/100 days, while around $t \approx +300$ days it increases up to 
$-0.8$ mag/100 days which is closer to the expected theoretical value of the Co-decay 
($-0.98$ mag/100 days). We deduce that SN~2004dj entered fully the radioactive 
nebular phase around $+350$ - $+400$ days after explosion. We analyse the bolometric light
curve further in Section 4 and 5.

\subsection{Spectroscopy}
\begin{table}
\begin{center}
\caption{Journal of spectroscopic observations. The columns contain the
following data: date and JD ($-$ 2450000) of the observations, days relative to the estimated
moment of explosion (JD 2453187, see Sect.4.1), central wavelength and airmass.}
\begin{tabular}{lcrcc}
\hline
Date & JD-2450000 & Days & $\lambda_c$ & Airmass \\
\hline
2004-08-17 & 3234.9 &47 & 5800& 1.57 \\
2004-08-20 & 3237.6 &50 & 5800& 2.92 \\
2004-08-22 & 3239.9 &52 & 5800& 1.47 \\
2004-08-25 & 3242.9 &55 & 5800& 1.49 \\
2004-09-03 & 3251.8 &64 & 5800& 1.49 \\
2004-09-04 & 3252.8 &65 & 5800& 1.53 \\
2004-09-06 & 3254.9 &67 & 5800& 1.04 \\
2004-09-22 & 3270.9 &83 & 5800& 1.22 \\
2004-10-02 & 3281.7 &94 & 5800& 1.24 \\
2004-10-03 & 3282.9 &95 & 5800& 2.02 \\
2004-10-05 & 3284.9 &97 & 5800& 2.25 \\
2004-10-06 & 3285.5 &98 & 5800& 1.13 \\
2004-10-07 & 3286.9 &99 & 5800& 1.70 \\
2004-11-14 & 3324.7 &137 & 5800& 1.32 \\
\hline
\end{tabular}
\end{center}
\end{table}

The spectroscopic observations were conducted at David Dunlap Observatory (DDO) with the
74" telescope and the Cassegrain spectrograph. The 100 lines/mm grating was applied 
giving a 2-pixel-resolution of about 800 at 6000 \AA. The journal of 
spectroscopic observations is presented in Table~6.  

The spectra were reduced and calibrated in the usual way using {\it IRAF}. After bias, flatfield and
cosmic-ray corrections, the spectrum was extracted with the {\it noao.twodspec.apall} task. The wavelength
calibration was done using two exposures of a FeNe spectral lamp made before and after each target 
observation. For flux calibration, HD~42818 (sp.type A0V, \citet{adelman}; \citet{kharit}) 
was observed as a flux standard. 

The DDO Cassegrain spectrgraph has a known limitation in obtaining low resolution spectra, because
the slit cannot be rotated, therefore the observations cannot be made at the parallactic angle.
Due to this technical limitation, the continuum slope of our spectra is certainly affected by
the differential atmospheric refraction, especially in the blue. The most distorted spectrum is
the one that was taken on 20th August, 2004, when the airmass was nearly 3 (see Table 5). On
the other hand, those spectra that were observed between airmasses of 1 and 1.5 are probably
less distorted, thus, they may be used for quantitative analysis. Nevertheless, the blue end of
all spectra was set as 4300 \AA~ in order to eliminate the most unreliable blue part.

In the following subsections we present and discuss the plateau and nebular spectra of SN~2004dj
separately. 

\subsubsection{Plateau phase}
\begin{figure}
\begin{center}
\psfig{file=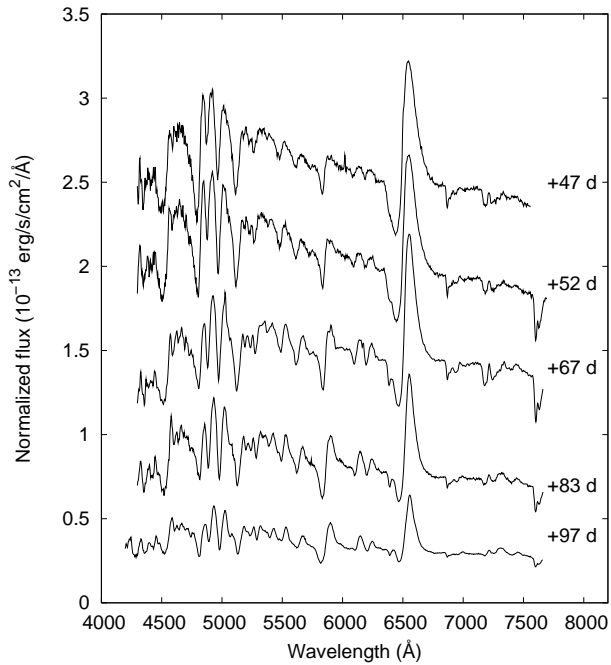,width=8.5cm}
\caption{Spectral evolution of SN 2004dj during the plateau and the transition phase.
Labels indicate the elapsed time since the estimated date of explosion in days.
The spectra have been shifted vertically by $5 \cdot 10^{-14}$ erg/s/cm$^2$/\AA~ 
for better visibility.}
\end{center}
\end{figure}

Fig.~9 shows five 
representative spectra from the 13 ones obtained during the plateau phase.
The plateau-phase spectra are typical for Type IIp SNe: dominant hydrogen lines with 
P Cygni profiles and moderately strong metallic line blends in the blue, 
superimposed on a continuum that is increasing toward the blue. Since
the spectral evolution is slow in this phase, the other spectra are very similar
to those plotted in Fig.~9. 

The evolution of the spectral features is also similar to other SNe Type IIp. 
When the ejecta cools, the continuum gets fainter and redder. The emission
component of $H \alpha$ becomes stronger, while the strength of $H \beta$
and the other excited Balmer lines decreases. Among the metallic lines, Na D
develops a strong P Cygni profile at the end of the plateau phase. Unfortunately,
no spectrum  earlier than $+47$ day was available for us, so we could not
check the presence of He in the early phases.
\begin{figure}
\begin{center}
\psfig{file=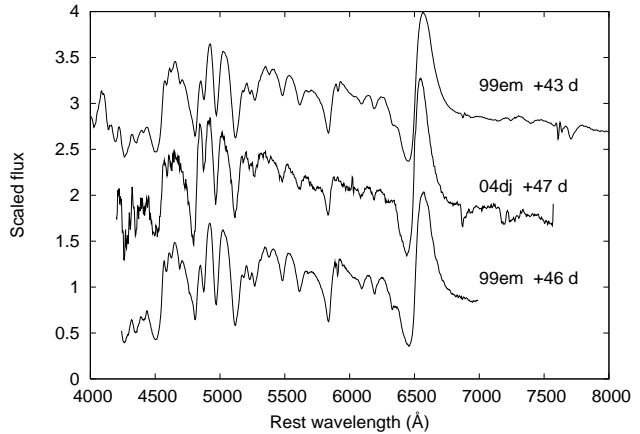,width=8.5cm}
\caption{Comparison of the spectra of SN~2004dj (thick line) and SN~1999em (thin line).
The time since explosion is labelled. The spectra were dereddened, corrected for
redshift of the host galaxy and shifted vertically for better visibility.}
\end{center}
\end{figure}

Our first spectrum of SN~2004dj was compared with the spectra of SN~1999em \citep{leonard1} 
made at similar phases. These comparison spectra were downloaded from the SUSPECT 
\footnote{http://bruford.nhn.ou.edu/$\sim$suspect/index1.html} database.  
Fig.~10 shows that the spectra of the two SNe are very similar. This is slightly
in contrast with the difference between the light curves of the two SNe found in Sect.2.1.1.

\begin{figure}
\begin{center}
\psfig{file=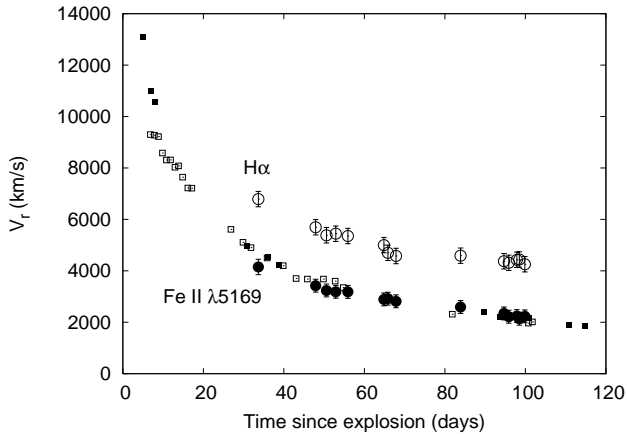,width=8.5cm}
\caption{Radial velocities of SN 2004dj. The photospheric velocities (filled circles) have been
computed from the Doppler-shift of the $\lambda 5169$ Fe~II line (see text). 
For comparison, photospheric velocities of SN~1999em (open squares, \citet{leonard1})
and SN~1999gi (filled squares, \citet{leonard2}) are also plotted. 
The $H\alpha$ velocities of SN~2004dj are shown as open circles. }
\end{center}
\end{figure}

From the photospheric spectra the radial velocities of the expanding ejecta were estimated
from the Doppler-shift of the $\lambda 5169$ \AA~ Fe~II line. Recently \citet{dh1} pointed
out that this line is a good indicator of the photospheric velocity. This line is also
favoured by \citet{hp} for estimating the expansion velocity in the middle of the plateau
phase. The velocities are plotted in Fig.~11 as filled circles. Because our spectra
do not cover the plateau phase well, the first radial velocity point was computed
from the line positions reported by \citet{patat1}. 
Fig.~11 also contains the radial velocities estimated from
the minimum of $H \alpha$. As it is expected, the optically thick $H\alpha$ forms
at higher atmospheric levels, probing higher ejecta velocities than the photospheric
Fe~II lines. 

\begin{table}
\begin{center}
\caption{Photospheric radial velocities of SN~2004dj corrected for redshift (+131 km/s) and 
host galaxy rotation (+90 km/s). The third column lists the average radial 
velocities calculated from the $\lambda 5169$ Fe~II line. The fourth column contains the velocities
based on the minimum of $H\alpha$. Errors are given in parentheses.}
\begin{tabular}{lrlc}
\hline
JD-2450000 & Days & $v_r$ (km/s) & $v_{H\alpha}$ (km/s) \\
\hline
3220.67$^1$ &33 & 4150 (300) & 6790 (300) \\
3234.89 &47 & 3423 (250)  & 5693 (300) \\ 
3237.59 &50 & 3236 (250) & 5382 (300) \\ 
3239.89 &52 & 3184 (250)  & 5444 (300) \\ 
3242.89 &55 & 3183 (250) & 5353 (300) \\ 
3251.86 &64 & 2888 (250)  & 5001 (300) \\ 
3252.85 &65 & 2917 (250)  & 4700 (300) \\ 
3254.86 &67 & 2813 (250)  & 4577 (300) \\ 
3270.90 &83 & 2592 (250)  & 4589 (300) \\ 
3281.67 &94 & 2342 (250)  & 4373 (300) \\ 
3282.87 &95 & 2214 (250) & 4312 (300) \\ 
3284.87 &97 & 2239 (250) & 4428 (300) \\ 
3285.49 &98 & 2136 (250) & 4443 (300) \\ 
3286.86 &99 & 2230 (250) & 4256 (300) \\ 
\hline	
\end{tabular}
\end{center}
$^1$ from \citet{patat1} \\
\end{table}

The measured radial velocities must be corrected for the motion of the host galaxy and
the SN itself within the host (the barycentric motion of the Earth is usually negligible
in the case of SNe). This is often done using the redshift of the host galaxy adopted
from major databases. In the case of SN~2004dj, the host galaxy, NGC 2403,
has a recession velocity of $+131$ km/s (NED). Moreover, the SN is located close the outer end
of a spiral arm, thus, its radial velocity is expected to differ from the systemic
velocity of the whole galaxy. Indeed, \citet{frater} recently mapped the kinematics of
NGC~2403 via H~I emission. It is visible in their Fig.~1 that the area around SN~2004dj
is on the receding side of the galaxy. The excess velocity was estimated to be $+90$ km/s,
thus, the total radial velocity of the barycenter of SN~2004dj was set as $+221$ km/s.
The final radial velocities of the SN ejecta were calculated with respect to this value
(see Table~7).

\subsubsection{Nebular phase}
\begin{figure}
\begin{center}
\psfig{file=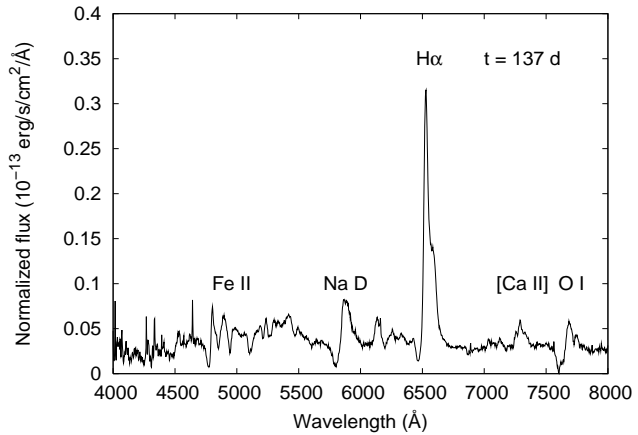,width=8.5cm}
\caption{A nebular spectrum of SN 2004dj, +137 days after the explosion. }
\end{center}
\end{figure}

Fig.~12 shows a single spectrum obtained at the beginning of the nebular phase ($t = +137$ days
after explosion). The spectrum is dominated by strong $H\alpha$ emission and P Cygni profiles
of Na~I, O~I and Fe~II. The forbidden $[$Ca~II$]$ line is also apparent around 7300 \AA. 
Recently \citet{chugai} analysed a more extensive set of nebular spectra of SN~2004dj.
They have pointed out significant asymmetry of the $H\alpha$ profile, from which they concluded
that the ejection of $^{56}$Ni (which is the primary source of $H\alpha$ emission in the
nebular phase via radioactive decay) was bipolar inside a spherically symmetric
envelope. We do not consider this topic in detail, except noting that our nebular spectrum 
of SN~2004dj (Fig.~12) do confirm the strong blueward asymmetry of the $H\alpha$ emission profile
found by \citet{chugai}. So far this phenomenon was reported only in two other SNe, 
namely SN~1987A \citep{pw} and SN~1999em \citep{elmha2}.

\section{Spectral modelling}

The photospheric spectrum obtained on Aug.17, 2004 (day $+47$, Table~6) is modelled using the
{\it SYNOW} parametrized spectrum synthesis code \citep{fisher,baron1,branch1,branch2}. 
This code assumes homologous expansion of the SN ejecta, i.e. $v(r) = r \cdot v_{ph}/R_{ph}$, where 
$r$ is the radius of a thin shell above the photosphere, $v(r)$ is the velocity of the shell,
$R_{ph}$ and $v_{ph}$ is the photospheric radius and the velocity at the photosphere, respectively.
Line formation is treated as in the Schuster-Schwarzschild model: the photosphere radiates
as a blackbody with temperature $T_{bb}$, and the lines are formed entirely in the envelope 
above the photosphere. The line formation is assumed due to pure scattering. 
This is equivalent to setting the line source function as 
\begin{equation}
S(r) ~=~ {I_{ph} \over 2} ~\left [ 1 - \sqrt{1 - \left ( {R_{ph} \over r} \right )^2} \right ]
\end{equation}
where $I_{ph} ~=~ B_{\lambda}(T_{bb})$ is the specific intensity at the photosphere.
The optical depth as a function of radius was expressed as a power law:
\begin{equation}
\tau(r) ~=~ \tau_{0} ~\left ( {r \over R_{ph}} \right )^{-\alpha}
\end{equation} 

In {\it SYNOW} the fit of the spectral lines is controlled by adjusting the optical 
depth of a reference line (a strong line in the optical regime) for each element. 
The depths of other lines are computed via the Boltzmann-equation. 
The solution of the radiative transfer equation is 
performed in the Sobolev approximation \citep{kasen}.  

Table~8 lists the elements identified in the day $+47$ spectrum, together with their
atomic parameters and fitted reference line optical depths. The model spectrum 
and the observed one are plotted in Fig.~13. The photospheric temperature was
set as $T_{bb} = 8000$ K determined from the slope of the red continuum. 
The velocity at the photosphere was chosen as $v_{ph} = 3400$ kms$^{-1}$, which
gives $R_{ph} = 922$ Mkm ($1324 ~ R_\odot$) for the radius of the photosphere 
$+47$ days after explosion. The optical depth exponent that gives the
best description of the observed spectrum was found as $\alpha = 6$.

The elements that could be
identified unambigously are H, Na~I, Sc~II and Fe~II. Ti~II is assumed to be responsible
for the blends below 4800 \AA, but other elements (e.g. Ca, Mg, Ni, Co) may
also contribute. Ba~II is also present, but its contribution is
weak and blended by Fe~II. Because the overall appearance of the spectrum is 
very similar to that of SN~1999em (Fig.~10), 
the more extensive line identification by \citet{leonard1} 
is probably also valid in this case. Fig.~14 shows the contribution of the
elements listed in Table~8 to the spectrum of SN~2004dj.

\begin{table}
\begin{center}
\caption{Reference line parameters and optical depths of the SYNOW model spectrum shown in Fig.~11.
The model has been computed assuming $T_{bb} = T_{exc} = 8000$ K, $v_{ph} = 3400$ kms$^{-1}$ and 
$\alpha = 6$ (see text). 
The columns give the following data: name and ionization state of the element, wavelength (in \AA),
oscillator strength, excitation potential (in eV) and the fitted optical depth 
for the reference line.}
\begin{tabular}{lcccc}
\hline
Ion & $\lambda$ (\AA) & log(gf) & $\chi$ (eV) & $\tau_{ref}$\\
\hline
H I & 6563 & 0.71 & 10.21 & 200 \\
Na I & 5890 & 0.12 & 0.00 & 1.1 \\
Sc II & 4247 & 0.32 & 0.32 & 1.0 \\
Ti II & 4550 & -0.45 & 1.58 & 8.0 \\
Fe II & 5018 & -1.40 & 2.89 & 4.0 \\
Ba II & 4554 & 0.19  & 0.00 & 2.0 \\
\hline
\end{tabular}
\end{center}
\end{table}

\begin{figure}
\begin{center}
\psfig{file=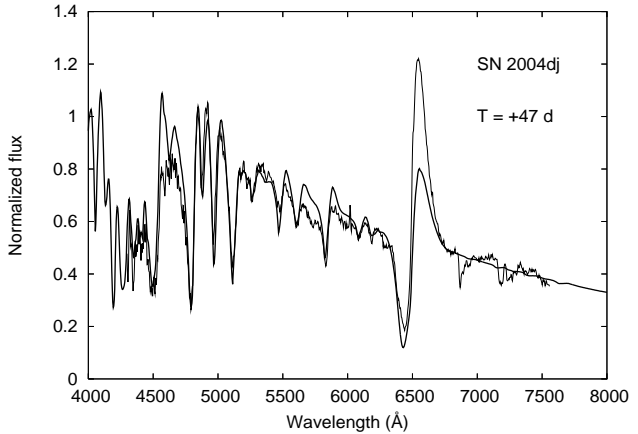,width=8.5cm}
\caption{The SYNOW model fit (thick line) and the $+47$ days observed spectrum (thin line). 
The model parameters are listed in Table 8. }
\end{center}
\end{figure}

\begin{figure}
\begin{center}
\psfig{file=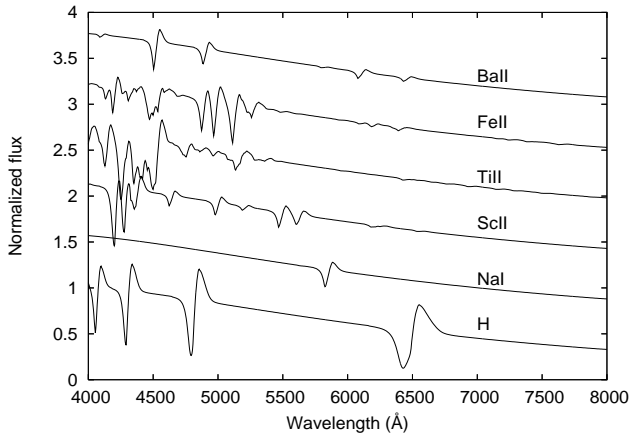,width=8.5cm}
\caption{The contribution of chemical elements to the observed spectrum. The labels on the right-hand
side indicate the composition. The adopted parameters are the same as in Table 8.}
\end{center}
\end{figure}

\subsection{The $H\alpha$ profile}
It is seen in Fig.~13 that $H\alpha$ cannot be fitted well by {\it SYNOW}: the absorption
is too deep and the emission is too low in the model spectrum. This is a usual situation
in modelling a strong line with a pure scattering source function \citep{branch1}. It does 
not necessarily mean the breakdown of the Sobolev approximation, but the source function
probably differs from that in Eq.1. 
\begin{figure}
\begin{center}
\psfig{file=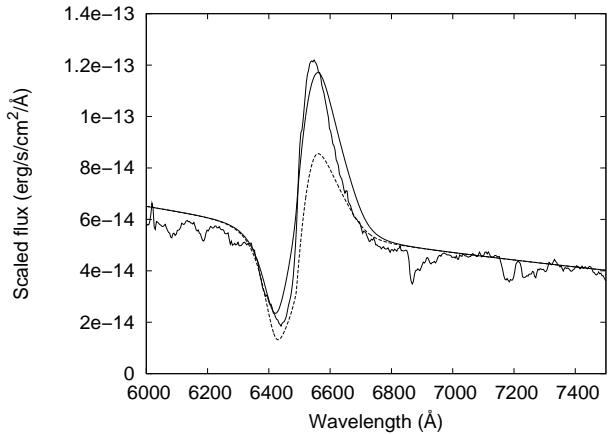,width=8.5cm}
\caption{The $H\alpha$ line profile computed with the scaled source function (Eq.4).
The continuous line corresponds to $K=2$, while the dashed line is for $K=1$. 
The observed line profile is also shown for comparison (thick line).}
\end{center}
\end{figure}

We found that a simple modification of the line source function can sufficiently explain
the emission content of $H\alpha$ within the Sobolev approximation. Now the source 
function takes the form
\begin{equation}
S(r) = K ~S_0(r)
\end{equation}
where $S_0(r)$ is the function given in Eq.1 and $K$ is a constant in $r$. In fact,
$K$ simply scales the photospheric intensity, but the shape of the source function
remains the same as the pure scattering one. Fig.~15 shows the line
profiles computed with $K=1$ and $2$. The agreement with the observed spectrum 
is much better when $K=2$, i.e. the photospheric intensity has an excess with
respect to the blackbody intensity at the wavelength of $H\alpha$. Since the
photosphere is actually a hydrogen recombination front in SNe IIp, the excess
emission can be interpreted as due to the recombination of hydrogen to the
$n = 3$ excited state followed by a downward transition to the $n=2$ state,
producing spontaneous emission at the wavelength of $H \alpha$. 
The exact form of the source function in SNe IIp atmospheres would deserve further study.
\citet{kasen} presented a technique for determining the source function (as well
as the optical depth) as a function of $r$. We leave this interesting problem
for future studies. 

\section{Distance determination}

There are a number of methods for determining SNe IIp distances. 
The Expanding Photosphere Method (EPM) \citep{kikw, hamuy, leonard1, leonard2, dh1}, 
a variant of the famous
Baade-Wesselink method, derives the angular diameter from both photometry
and radial velocities, and their comparison yields the distance. 
The more sophisticated extension of this method is the 
Spectral Fitting Expanding Atmosphere Method (SEAM) \citep{bhb}, which uses
a custom-crafted SN model atmosphere and full treatment of hydrodynamics
and radiative transfer to fit model spectra to the observed ones.  SEAM
has been succesfully applied to SNe IIp and other types as well. SEAM stands on
more solid physical basis than EPM, but requires much more computing power.
  
The Standard Candle Method (SCM) \citep{hp} relies on the correlation between
the absolute magnitude and the radial velocity of the SN ejecta in the middle
of the plateau phase. Here the distance can be found by comparing the observed
and the calculated absolute magnitudes, like in the case of Cepheids. The so-called
"Plateau-Tail Method" (PTM) \citep{nady} compares the SN light curve with 
theoretical models and derive the distance via the model parameters. 

In the following we apply EPM and SCM to derive the distance to SN~2004dj,
and compare it with distances of the host galaxy found by other methods. 

\subsection{Expanding Photosphere Method}

The methodology of EPM has been recently reviewed and thoroughly
discussed by \citet{dh1}. They concluded that EPM is best used
at early phases, when the SN ejecta is fully ionized and continuum
opacity dominates line opacities in the optical. Unfortunately,
SN~2004dj could not be observed at such early phases, which makes
the application of EPM problematic. Nevertheless, we try to use
EPM for the earliest observed data of SN~2004dj and critically
compare the results with other distance estimates to minimize
the systematic errors caused by the possible failure of the
asssumptions of EPM.

EPM has been applied in the same way as described by \citet{vinko1}.
The angular radius has been derived from the $UVOIR$ bolometric
light curve (Sect.2.1.3) as
\begin{equation}
\theta ~=~ \sqrt{ {f_{bol} \over {\xi^2(T) \sigma T_{eff}^4} } }
\end{equation}
where $f_{bol}$ is the "observed" bolometric flux, $T_{eff}$ is 
an effective temperature of the diluted blackbody and $\xi(T)$
is the dilution factor (determined from SNe IIp model 
atmospheres). For each epoch, the effective temperature
has been estimated by fitting blackbodies to the dereddened $BVI$ and
$VI$ fluxes (the $R$ band has been omitted because of the presence
of $H\alpha$ emission). 

The dilution factor $\xi(T)$ in Eq.5 comes from model atmospheres.
We have adopted the dilution factors derived by \citet{eastman}
and parametrized by \citet{hamuy}. Recently \citet{dh1} derived
new dilution factors that are systematically higher than those
of \citet{eastman}. The effect of using these dilution factors
on the distance will be discussed later in this Section.
Note that the dilution factors depend on the bands used for 
temperature determination, thus, they are slightly different 
for $T_{BVI}$ and $T_{VI}$.

Assuming spherically symmetric SN ejecta and homologous expansion, 
for each oberved epoch $t$, the distance $D$ is given by the linear equation
\begin{equation}
t ~=~ t_0 ~+~ D {\theta \over v_{ph}} 
\end{equation}
where $t_0$ is the moment of explosion, $v_{ph}$ is the photospheric
expansion velocity at epoch $t$. The velocities have been interpolated
from the observed values listed in Table~7. The derived parameters 
relevant for the EPM analysis are listed in Table~9. 
\begin{table}
\begin{center}
\caption{Quantities derived during the EPM. The columns contain the followings:
JD $-2450000$ of observation, temperatures (in Kelvins), angular sizes (in $10^9$ km Mpc$^{-1}$),
and values of $\theta / v_{ph}$ (in day Mpc$^{-1}$).  }
\begin{tabular}{lrrllrl}
\hline
$t$ & $T_{BVI}$ & $T_{VI}$ & $\theta_{BVI}$ & $\theta_{VI}$ & $(\theta/v)_{BVI}$ &$(\theta/v)_{VI}$ \\
\hline
3223.4 & 7047 & 10602 & 3.347 & 1.700 &  9.674 & 4.912 \\  
3226.6 & 6632 &  9988 & 3.618 & 1.967 & 10.984 & 5.973 \\
3228.6 & 6419 &  9795 & 3.684 & 2.033 & 11.568 & 6.385 \\
3229.6 & 6412 &  9727 & 3.687 & 2.062 & 11.752 & 6.571 \\
3230.4 & 6823 &  9851 & 3.570 & 2.043 & 11.511 & 6.589 \\
3234.4 & 5955 &  9653 & 3.722 & 2.023 & 12.841 & 6.980 \\
3236.6 & 6079 &  9385 & 3.729 & 2.163 & 13.340 & 7.740 \\
\hline
\end{tabular}
\end{center}
\end{table}   

\begin{figure}
\begin{center}
\psfig{file=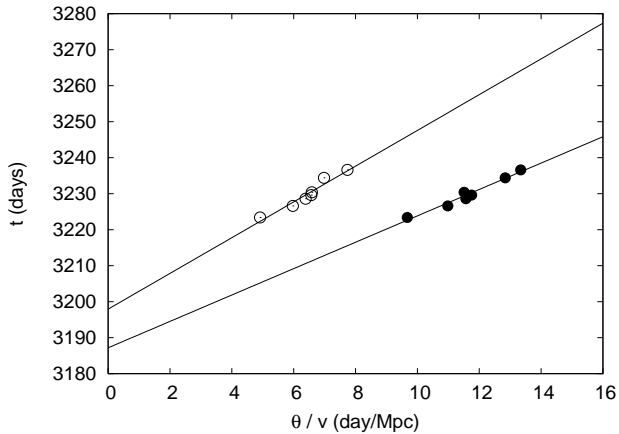,width=8.5cm}
\caption{Distance determination in EPM via Eq.6. Parameters from $T_{BVI}$ are plotted
with filled circles, while those from $T_{VI}$ are represented by open symbols. The fitted
lines give the distance and the moment of explosion (see text).}
\end{center}
\end{figure}

\begin{table}
\begin{center}
\caption{Results of EPM analysis. The uncertainties are given in parentheses. The quality
of the fit (rms) is also shown.}
\begin{tabular}{cccc}
\hline
Bands & $t_0$  & $D$  & rms \\
 & (JD-2450000) & (Mpc) & \\
\hline
$BVI$ & 3187 (3) & 3.66 (0.3) & 0.77 \\ 
$VI$ & 3198 (4) & 4.97 (0.6) & 1.14 \\
\hline
\end{tabular}
\end{center}
\end{table}

The EPM quantities are plotted in Fig.~16. The slope and the zero point of the fitted line 
gives the distance and the moment of explosion, respectively (see Eq.6). It is visible that
the data from $T_{BVI}$ and $T_{VI}$ give quite different parameters, which is a usual situation
in EPM \citep{leonard1, leonard2}. 

Which solution is closer to the real one? The concept of EPM (diluted blackbody, spherical 
symmetry, etc.) received many criticism recently \citep{leonard3, baron2}, leading to a
conclusion that EPM is at least "suspicious" and the results are systematically off from 
the real values. The discussion of all the problems is beyond the scope of this paper.
Here we would like to show that combining the EPM results with those of other methods
and the observations, it may be possible to select an EPM solution that is consistent with
other results.

\begin{figure}
\begin{center}
\psfig{file=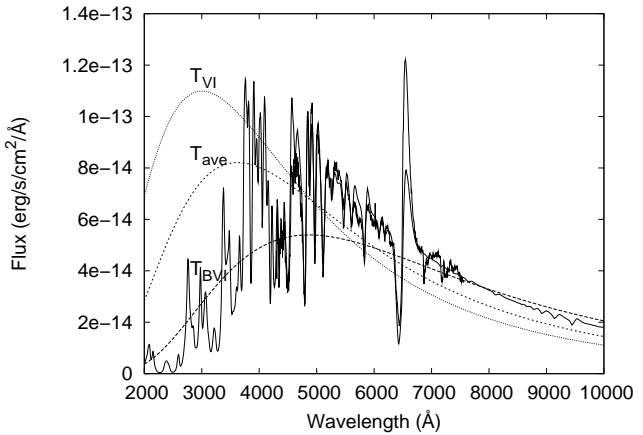,width=8.5cm}
\caption{Flux distribution of blackbodies fitted in EPM, together with the observed and
model spectrum on day $+47$ (Sect.3). The drop of flux in the blue region caused
by metallic line blends (Fig.~14) and the continuum slope in the red cannot be fitted
by a single blackbody. However, the total flux (but not its spectral distriution)
may be adequately described by the blackbody with $T_{BVI}$.}
\end{center}
\end{figure}

A critical step in EPM is the determination of the effective temperature, since it 
affects the calculated blackbody flux as well as the dilution factor.
This quantity strongly depends on the selected wavelength bands (Table~9). Recently 
\citet{dh1} and \citet{baron2}
showed that such kind of fitted blackbody temperature does not represent well the 
real spectral flux distribution of the SN atmosphere. Indeed, comparing the temperatures
in Table~9 and the result of spectral modelling in Sect.3, it is clear that $T_{BVI}$ is
too low, while $T_{VI}$ is too high with respect to the temperature ($T_{bb}=8000$ K) 
that represents the spectrum observed on day $+47$ well. This is illustrated
further in Fig.~17, where the observed spectrum and the {\it SYNOW} model computation are plotted
together with blackbody flux distributions corresponding to 
$T_{BVI} = 5955$ K, $T_{VI}=9653$ K and the average of these two temperatures. 
The blackbody curves were scaled according to the angular radius of the photosphere 
found by EPM.  

It is clear that the continuum slope on the red side of the observed spectrum 
is indicative of higher temperature, but a blackbody with such a temperature fails to
model the sudden drop of the flux below 4000 \AA. Similar figures have been presented
by \citet{dh1} and \citet{baron2} for SN~1999em.
There is no doubt that the blackbody
assumption is invalid, even in the very early phases (that were actually not observed in the
case of SN~2004dj). The cause of this disagreement has been already shown in Fig.~14: the 
metallic line blends (especially the Ti~II absorption) are responsible for
the sudden decrease of the flux in the blue. 

Fig.~17 shows that the blackbody with $T_{BVI}$ has a flux distribution
that does not give the best fit between 4000 and 6000 \AA, but it adeqautely describes both
the flux drop in the blue and the tail in the red. The two other curves that are closer to
the real photospheric temperature have large excess flux below 4000 \AA. These curves
highly overestimate the photospheric flux in the blue bands, increasing the EPM distance.
The $T_{BVI}$ blackbody, however, results in a {\it bolometric} flux that is closer to
the real bolometric luminosity, even though the spectral distribution is not 
perfectly matched. Since the purpose of EPM is to estimate the distance, rather than
to model the spectra, the diluted blackbody assumption may be an acceptable 
compromise, if one regards it as the representation of the amount of integrated flux 
of the photosphere, rather than a physically correct model of the spectral flux 
distribution.

In principle, the application of a proper dilution factor should correct the flux 
for both the temperature mismatch between the SN and the blackbody, and
the line blanketing. In this ideal case the same distance could be derived
from any filter combinations. However, this is true only if the model spectrum (that
was used to compute the dilution factor) perfectly fits the observed one.
In reality, the metallic lines in the blue band can vary significantly from
SN to SN, which can be accounted for only with custom-crafted model atmospheres.
This makes the usage of $B$ band fluxes and temperature such as $T_{BV}$ less 
effective for EPM. 

Recently \citet{dh1} presented a new set of dilution factors appropriate for
SNe IIp, that are systematically higher by 10 - 20 \% than those given by
\citet{eastman}. In a follow-up paper \citet{dh2} showed that the application
of their new dilution factors for SN~1999em increases its EPM-distance,
bringing it into good agreement with the Cepheid-distance of the host galaxy,
thus, eliminating most of the systematic errors of EPM found by \citet{leonard3}. 
If one applies the dilution factors of \citet{dh1} for the present data of 
SN~2004dj, its EPM-distance will increase as $D_{BVI} = 4.3 \pm 0.3$ and
$D_{VI} = 6.2 \pm 0.7$ Mpc. As in the case of SN~1999em, these distances
are also systematically higher than those that were derived from the 
Eastman et al. dilution factors. The problem is that for SN~2004dj the
EPM-distances are all {\it longer} than the values found by other methods 
(see below). In this case the application of the dilution factors of
\citet{dh1} even increases the discrepancy between EPM and the other methods.
It might be possible that the SNe IIp models by \citet{eastman} represents 
SN~2004dj better than SN~1999em, while the models applied by \citet{dh1}
are better for SN~1999em. On the other hand, regarding the similarity 
of the optical spectra  of these two SNe (Fig.10), it is hard to believe
that the atmospheres are so different. Detailed modelling of the
spectra of SN~2004dj may help to resolve this discrepancy. 
          
Concluding the EPM analysis, we adopt the solution based on the Eastman et al. 
dilution factors and $T_{BVI}$ (Table~10), namely $t_0 = 2453187 \pm 3$ and $D=3.66 \pm 0.3$ Mpc 
as the moment of explosion and distance, respectively. In the followings we show that
this distance is in agreement with other distance estimates for the host galaxy and
SN~2004dj itself.   

\subsection{Standard Candle Method}

The formulae for the SCM have been selected from \citet{hamuy1} as follows:
\begin{equation}
V_{50} - A_V + 6.564 \log({v_{50} \over 5000}) ~=~ 5 \log(H_0 \cdot D) -1.478 
\end{equation}
\begin{equation}
I_{50} - A_I + 5.869 \log({v_{50} \over 5000}) ~=~ 5 \log(H_0 \cdot D) -1.926
\end{equation}
where the observed quantities ($V_{50}$, $I_{50}$, $v_{50}$) are the $V$ and $I$ magnitudes
and the radial velocity (in km/s) 50 days past explosion (i.e. in the middle of the plateau phase). 
Ideally, the distances obtained from the $V$ and $I$ data should be the same, therefore
this method can also be used to test the consistency of the applied reddening correction. 

From Table~3 and Table~6 we have chosen $V_{50} = 12.04 \pm 0.03$, $I_{50} = 11.40 \pm 0.03$
and $v_{50} = 3250 \pm 250$ km/s. The absorption corrections were computed from $E(B-V) = 0.07$
mag and the galactic reddening law, as before. For the Hubble constant 
$H_0 = 73$ km/s/Mpc 
has been adopted \citep{riess}. This value is now based on both SNe~Ia and 
Cepheid variables, and brings the previously conflicting distance scales into agreement. 
From these data we obtained $D_V = 3.47 \pm 0.2$ Mpc and $D_I = 3.52 \pm 0.2$ Mpc from the
$V$ and $I$ calibration, respectively. The uncertainties are purely from the observational 
errors. These distances agree nicely with each other and with the EPM distance 
($D_{EPM} = 3.66 \pm 0.3$ Mpc) within the errors, strengthening the credibility 
of both of the distance and reddening estimates. 
  
\subsection{Comparison with other distance determinations}  

The distances computed above are all based on the photometric and spectroscopic
behaviour of the SN ejecta during the plateau phase. The overall uncertainty
of such distances is about 20 \% in the case of SCM \citep{hp}, while it can
be as large as 60 \% for EPM \citep{leonard3}. Thus, the SN-distances should
be compared with other distance estimates based on other methods, in order
to test the possible systematic errors. 

Since the host galaxy, NGC~2403, a member of the M81 group, is a nearby, 
well-studied object, its distance
has been estimated from many other methods. These are listed in Table~11
together with the SN-based distances.

The Cepheid distance \citep{freedman1} was usually applied in previous studies 
of SN~2004dj, but it is worth noting that this value is based on only $I$ 
photometry without direct reddening determination. \citet{freedman1} 
{\it assumed} $A_I = 0.2$ mag as the $I$-band absorption during the distance
measurement. From Table~11 it
is visible that the SN-distances are in agreement with the recent 
Tully-Fisher distance, and differ less than $2 \sigma$ from the Cepheid
distance. This is not a significant difference, if one takes into account
the additional uncertainty of the Cepheid distance due to the lack of reddening
information. We conclude that the average distance of NGC~2403, 
$$ D~=~ 3.47 \pm 0.29 ~{\rm Mpc} $$
is consistent with all available distance estimates within the errors. 

\begin{table}
\begin{center}
\caption{Comparison of distance estimates for SN~2004dj and NGC~2403.}
\begin{tabular}{lcl}
\hline
Method & $D$ (Mpc) & Reference \\
\hline
Tully-Fisher & $3.50 \pm 0.31$ & $LEDA$; \citet{russell} \\
Cepheid P-L & $3.22 \pm 0.15$ & \citet{freedman1} \\
EPM & $3.66 \pm 0.30$ & present paper \\
SCM & $3.50 \pm 0.20$ & present paper \\
\hline
average & $3.47 \pm 0.29$ & \\
\hline
\end{tabular}
\end{center}
\end{table}

\section{Physical parameters}

\subsection{Nickel mass}

\begin{figure}
\begin{center}
\psfig{file=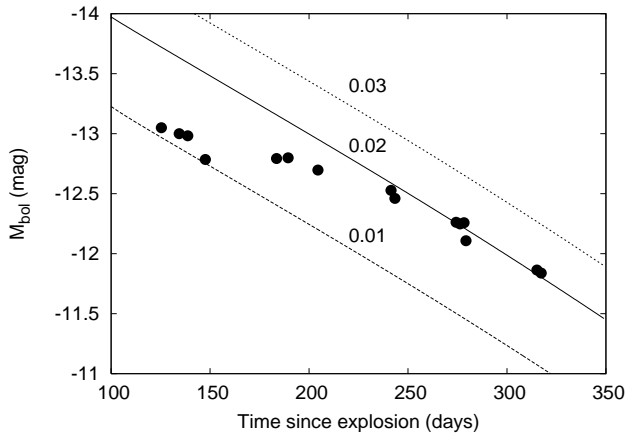,width=8.5cm}
\caption{Comparison of absolute bolometric magnitudes of SN 2004dj (filled circles) 
with model computations (lines) during the early tail phase. 
Labels indicate the Ni-masses (in $M_{\odot}$). }
\end{center}
\end{figure}

During the tail phase the light curve is thought to be powered by radioactive decay
of $^{56}$Co - $^{56}$Fe. Since $^{56}$Co is the daughter nucleus of $^{56}$Ni, the
luminosity is proportional to the initial nickel mass produced by the explosion.

In order to estimate the nickel mass of SN~2004dj we compared the absolute bolometric
light curve (calculated from the "observed" $UVOIR$ bolometric light curve 
discussed in Sect.2.1.3, and the distance estimated in Sect.4) with the prediction
of the simple light curve model described by \citet{vinko1}. This model calculates
the luminosity in a homologously expanding SN ejecta, assuming radioactive decay,
full gamma-ray and positron trapping and prompt thermalization of gamma photons and
positrons. The density structure of the ejecta has been given as a power-law with
exponent $-7$. This value is derived from the radius dependence of the optical
depth $\tau (r) ~=~ \tau_0 (r / R_{ph})^{-6}$ (found in Sect.4) by computing
$\rho (r) ~=~ - \kappa^{-1} (d \tau / d r)$ and assuming spatially constant $\kappa$.
The ejected mass has been set as $M_{ej} = 20 M_{\odot}$ (see Sect. 5.2). The calculated
light curves were mostly sensitive to the initial nickel mass, which was treated as
a free parameter. Fig.~18 shows the light curves (thin lines) calculated with
$M_{Ni} = 0.01, ~0.02$ and $0.03 ~M_{\odot}$. The observational data are also plotted
for comparison as filled circles. Although during the observations SN~2004dj did not
enter into the radioactive tail phase completely (the slope of the observed light curve
differs from the calculated one given by the Co-decay), it is seen that the luminosity
level can be explained with $M_{Ni} = 0.02 \pm 0.01 ~M_{\odot}$. This nickel mass is
the same as derived by \citet{chugai} from the comparison of the absolute $V$ light 
curves of SN~2004dj and SN~1987A. Based on the steepness of the $V$ light curve measured
after 100 days \citet{kotak} got $M_{Ni} \sim 0.022 ~M_{\odot}$. 

Alternatively, the nickel mass can be determined using semi-empirical correlations,
such as those presented by \citet{elmha1}. Using their Eq.2 one can compute the nickel
mass from the absolute $V$ magnitude 35 days before the light curve inflection point. 
The moment of inflection point was found as $T_i = 2453297$ JD (Sect.2.1.1), and the
absolute magnitude 35 days before that is 
$M_V (T_i - 35) = -15.52 \pm 0.1$ mag. The nickel mass is then 
$M_{Ni} = 0.022 \pm 0.002 ~M_{\odot}$, in perfect agreement with the value derived above.
From the steepness of the light curve during the transition phase \citet{chugai}
got $M_{Ni} = 0.013 \pm 0.004 M_{\odot}$. It is concluded that all the methods applied
here agree that the estimated nickel mass of SN~2004dj is 
$M_{Ni} = 0.02 \pm 0.01 ~M_{\odot}$. This is slightly below the nickel mass of SN~1999em and
SN~1999gi \citep{elmha1}, which is an expected result based on the light curves (Sect.2.1.1).

\subsection{Progenitor mass}

\begin{figure*}
\begin{center}
\leavevmode
\psfig{file=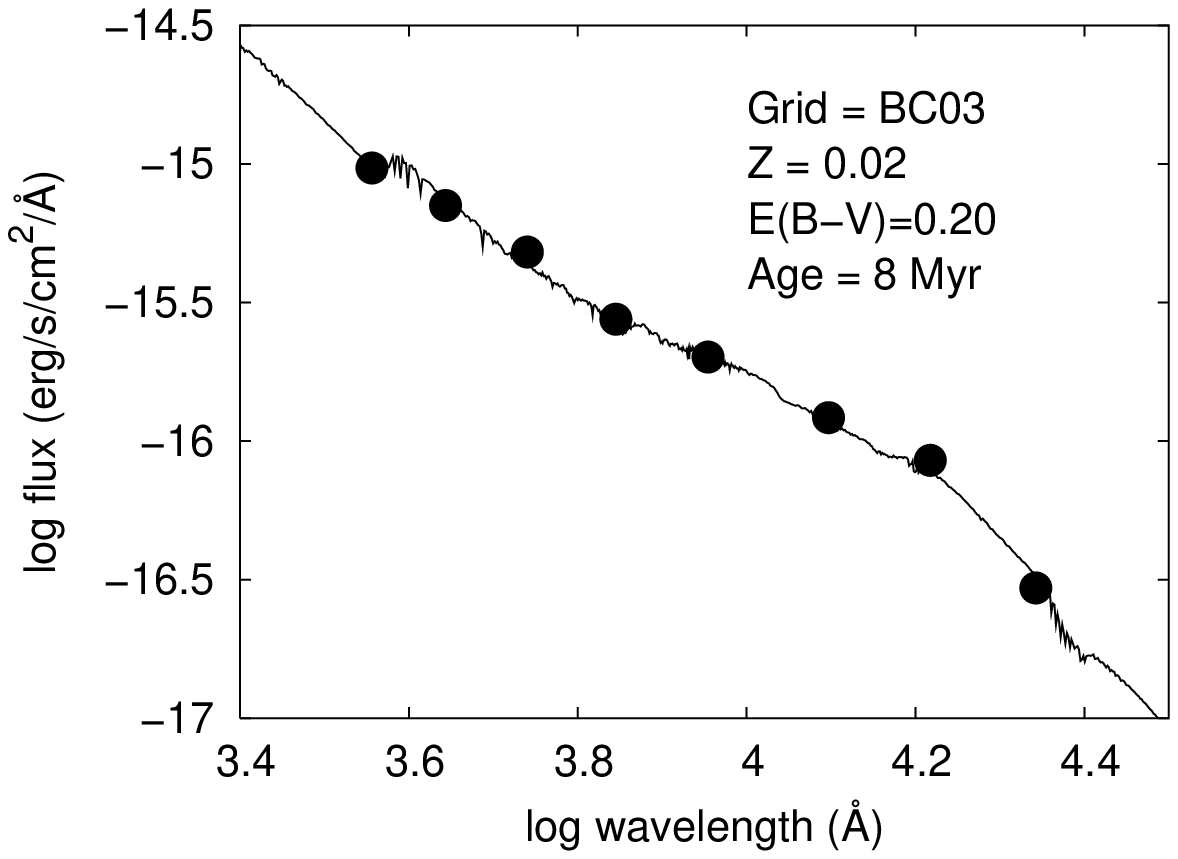,width=8cm}
\psfig{file=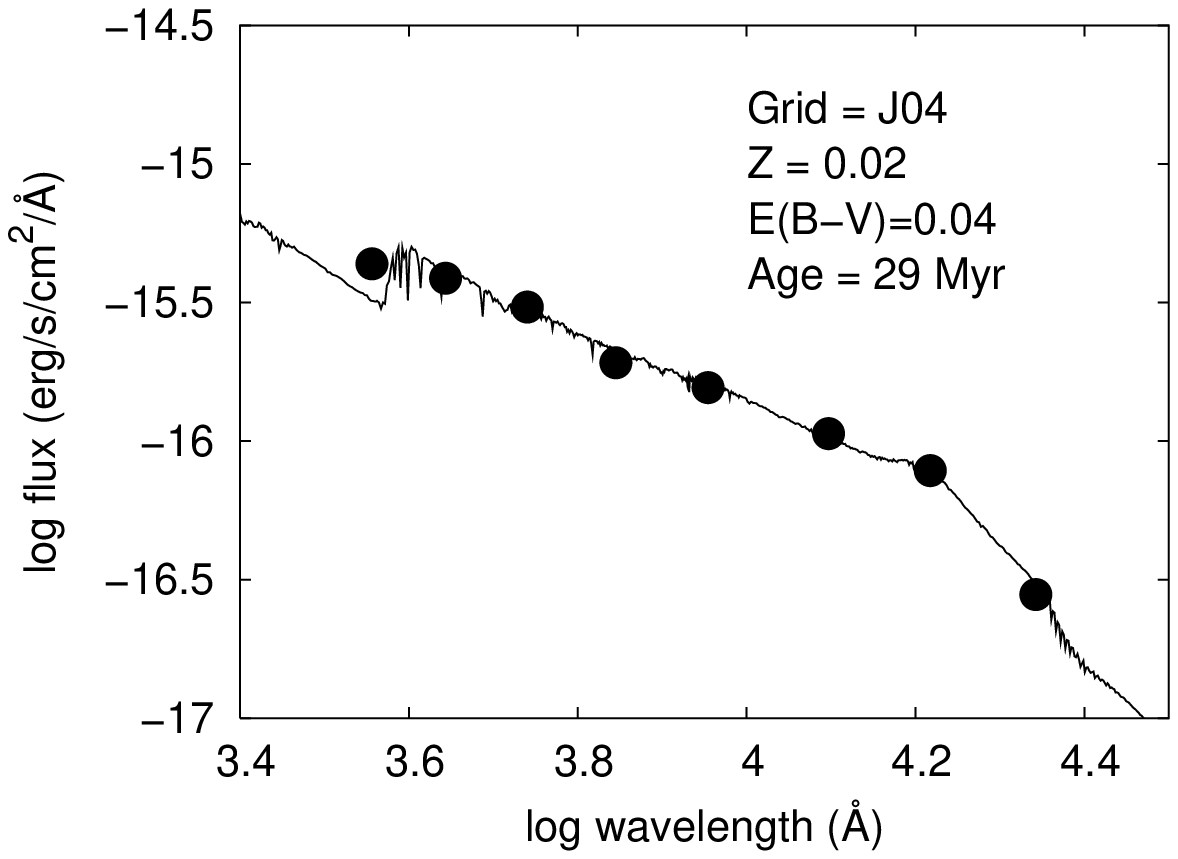,width=8cm}
\psfig{file=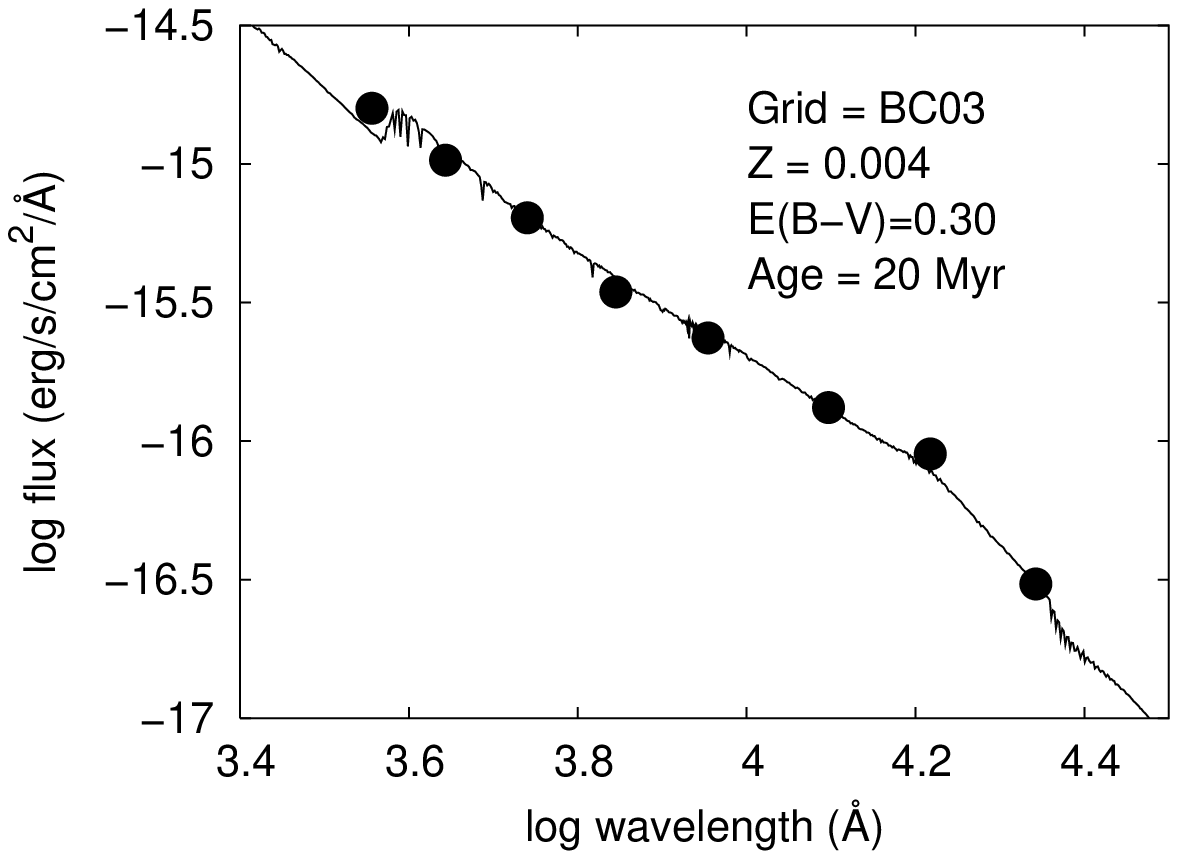,width=8cm}
\psfig{file=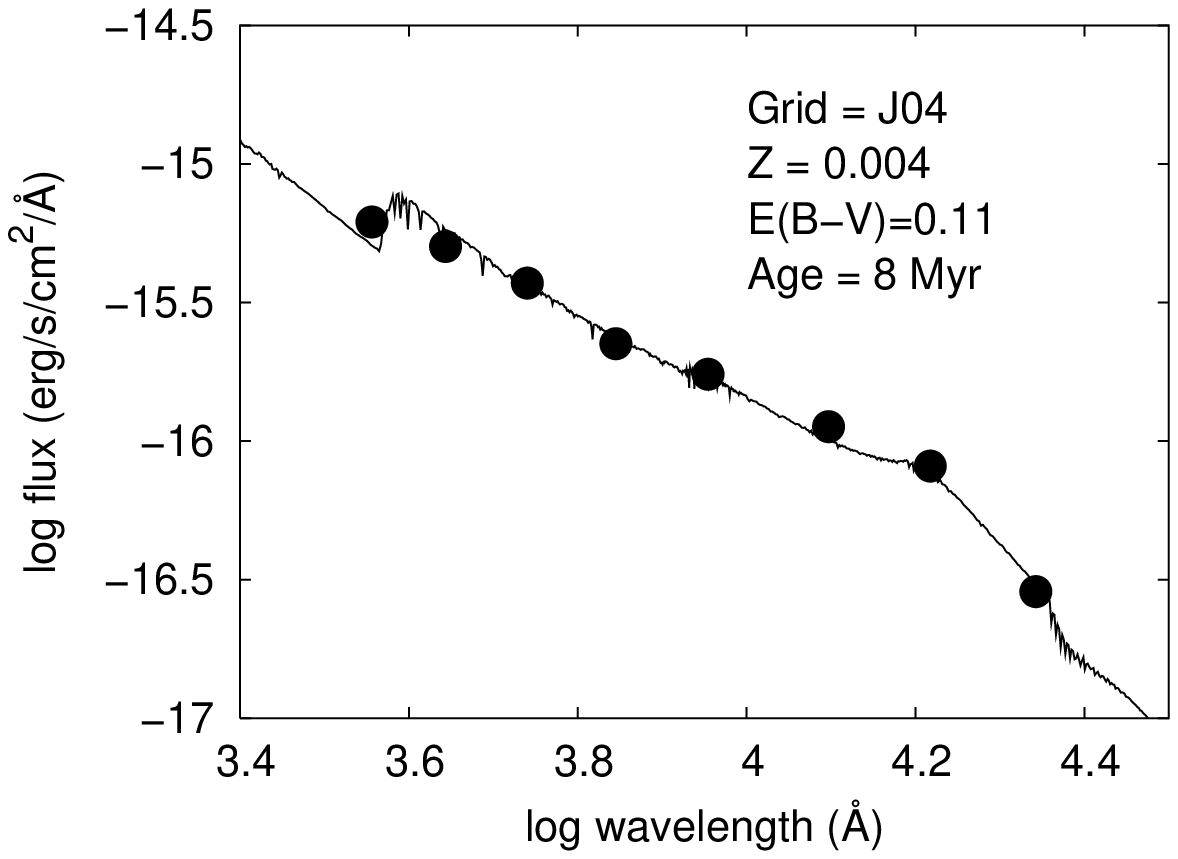,width=8cm}
\caption{Results of SED fitting to the observed data of S96. The applied grids 
and the fitting parameters are indicated on each panel. See Table~ 13 for more
information.}
\end{center}
\end{figure*}

Up to now the progenitors of Type II SNe have been identified directly
in a number of cases (see Sect.1 for references). Surprisingly, these turned out to
be relatively less massive stars, some of them being close to the 
theoretical limit of core collapse ($\sim 8 M_{\odot}$). 
However, in most cases the progenitor could be detected only
in a single bandpass, preventing the identification of its colour and  
evolutionary state. 

SN~2004dj is particularly interesting in this respect,
because its progenitor was a member of a compact cluster (Sandage~96) in
NGC~2403. The age of a cluster can be determined more precisely than that
of a single star, thus, the evolutionary state of the progenitor could be
secured. Two groups have reported their observations of S96, made some years 
before the explosion \citep{maiz,wang} and their analysis of the cluster 
SED (see Sect.1). 

We have re-analysed the SED of S96 based on our own $BVRI$ photometry
(Table~2) supplemented by data from the literature \citep{larsen,skrut}.
Our optical photometry is in very good agreement with the observations 
published by \citet{maiz}, except for the $B-V$ colour index, where
our value is closer to the one given by \citet{chugai}. The adopted
magnitudes of S96 are shown in Table~12. Note that \citet{maiz}
reported a Gunn-Johnson color index $u-B = 0.41 \pm 0.06$ that differs
from the $U-B=-0.51$ value  given by \citet{larsen} in the Johnson system.
We have preferred the latter value, since it is in the same photometric system
as the other optical fluxes, but the $U$ magnitude seems to be more uncertain
than the fluxes in other bands. It may affect the derived cluster age, since
it is connected with the amplitude of the Balmer-jump that increases with age.  

\begin{table}
\begin{center}
\caption{The adopted magnitudes for Sandage 96 (see Section 2.1 and Table~2).}
\begin{tabular}{cccl}
\hline
Filter & $\lambda$ (\AA) & Mag. & Reference\\
\hline
U & 3600 & 17.68 & \citet{larsen} \\ 
B & 4400 & 18.26 & present paper \\
V & 5500 & 17.85 & present paper \\
R & 7000 & 17.53 & present paper \\
I & 9000 & 17.06 & present paper \\
J & 12500 & 16.19 & \citet{skrut} \\
H & 16500 & 15.54 & \citet{skrut} \\
K & 22000 & 15.42 & \citet{skrut} \\
\hline
\end{tabular}
\end{center}
\end{table}

We have selected two recent grids for calculating the model SEDs: the widely used model
sequence by \citet{bc03} (BC03), based on Padova evolutionary tracks 
(that was also applied for S96 by \citet{wang}), and a more recent
one published by \citet{jim} (J04), containing new stellar interior models, evolutionary tracks
and improved treatment of mass loss. 
For the cluster initial mass function, a Salpeter IMF was adopted. It is known that 
the Salpeter function does not describe well the shape of the IMF, especially in the
low mass regime. The selection of the Salpeter IMF was motivated mainly by the fact that
the two previous studies for S96 also used this function, and we wanted to make the 
comparison of different solutions easier. 
For the same reason, a single starburst was assumed as the star formation history of S96.

The fitting has been computed via a simple $\chi^2$ minimization. The observed magnitudes
(Table~12) were transformed into monochromatic fluxes (in erg/s/cm$^2$/\AA) using the calibration
by \citet{hamuy} (see Sect. 2.1.3), dereddened, 
and compared with the predictions of the spectral synthesis model grids. 
The fitted parameters were the cluster age $T_{cl}$, the cluster mass 
$M_{cl}$ and the overall cluster reddening $E(B-V)_{cl}$. The metallicity of the cluster
was kept fixed during the minimization. It should be noted that 
$E(B-V)_{cl}$ may be quite different from the reddening derived for SN~2004dj, because
the compact cluster may contain a significant amount of intracluster gas and dust 
that is situated behind the SN. 
On the other hand, these parameters are correlated, and their simultaneous fitting may 
lead to spurious solutions. This is especially true for the reddening and the age,
because a higher reddening correction would result in a bluer SED, indicating a lower 
age.

\begin{table*}
\begin{center}
\caption{Results of SED fitting for S96. The columns contain the followings:
Model sequence, metallicity, cluster age (in Myr), cluster mass 
(in $10^3 ~M_{\odot}$), 
cluster reddening (in mag), predicted SN mass for the given cluster age 
(in $M_{\odot}$, estimated from Padova isochrones), reduced $\chi^2$ and reference. 
The last three rows list the solutions published by \citet{maiz} and \citet{wang}.}
\begin{tabular}{cccccccl}
\hline
Grid & $Z$ & $T_{cl}$ & $M_{cl}$ & $E(B-V)_{cl}$ & $M_{SN}$ & $\chi^2$ & reference \\
  &  &  (Myr) & ($10^3 ~ M_{\odot}$) & (mag) & ($M_{\odot}$) &  & \\
\hline
BC03 & 0.02 & 8 & 38 & 0.20 & 24 & 0.47 & present paper\\
J04 & 0.02 & 29 & 107 & 0.04 & 9.5 & 1.20 & present paper \\
BC03 & 0.004 & 20 & 141 & 0.30 & 12 & 0.95 & present paper \\
J04 & 0.004 & 8 & 29 & 0.11 & 26 & 0.96 & present paper \\
\hline
SB99 & 0.02 & 14 & 24 & 0.13 & 15 & 0.16 & \citet{maiz} \\
SB99 & 0.02 & 29 & 57 & 0.23 & 9.5 & 1.58 & \citet{maiz} \\
BC03 & 0.008 & 20 & 96 & 0.35 & 12.5 & 0.77 & \citet{wang} \\ 
\hline     
\end{tabular}
\end{center}
\end{table*}

The resulting parameters corresponding to the minimum of $\chi^2$ 
are listed in Table~13 and plotted in Fig.~19. Because the metallicity
of NGC 2403 at the position of S96 is probably subsolar ($[$O/H$] = -0.24$, 
\citet{pily,wang}), we have used the models with $Z=0.02$ and $Z=0.004$ 
metallicities from both grids. Comparing these results with the published
ones (shown in the last three rows of Table~13), it is visible that our analysis 
resulted in the same kinds of solutions as the previous studies: 
a "young" population with $T_{cl} = 8$ Myr and an
"old" population with $20 < T_{cl} < 30$ Myr. The "old" solutions are
very similar to the ones given in the previous studies. The "young" solution,
however, has considerably less age. It is interesting that this solution 
comes out using both model grids, regardless of metallicity (but with different
reddenings and cluster masses, of course).  
Based on the $\chi^2$ values,
the "young" solution from the BC03 grid gives the best description of
the observed data. However, we agree with \citet{wang} that if the observed
SED is undersampled, such as the one presented here, it may be misleading
to accept or reject solutions based on $\chi^2$ values only. Thus, we conclude
that, at present, there are three probable solutions for the SED fitting of
S96: the old one corresponds to $20 < T_{cl} <30$ Myr, the young one with
$T_{cl} \approx 8$ Myr, and a solution with age between these two, 
$T_{cl} = 14$ Myr. Each solutions have their own combinations of cluster 
reddening and total mass, between the values of $0.04 < E(B-V) = 0.35$ and
$30 < M_{cl} < 140 \cdot 10^3 ~M_{\odot}$. Due to the correlations between
these parameters, it is not possible to assign a unique
solution for all these quantities simultaneously. 

Regarding the potential progenitor of SN~2004dj, it is seen in Table~13 that
the SN mass (determined from the Padova isochrones, \citet{girardi}) 
ranges from $\sim 10 ~M_{\odot}$ to $\sim 25 ~M_{\odot}$, depending on
age. The uncertainty of the
age determined from SED fitting is at least 10 \% \citep{jim}, implying an error
of the SN-mass of about $\pm 4 ~M_{\odot}$ for the young- 
and $\pm 1 ~M_{\odot}$ for the old solution.  
It is interesting that the young (8 Myr) solution results in significantly
higher mass than all the SNe IIp progenitor masses detected so far. 
The highest mass of a detected SN Type II progenitor is that of SN~1999ev 
\citep{maund1} with $15 ~M_{\odot} < M_{ZAMS} < 20~M_{\odot}$. This is
similar to the progenitor mass of SN~2004et, being 
$M_{ZAMS} = 15^{+5}_{-2} ~M_{\odot}$ \citep{li1} and SN~1999gi
with $M_{ZAMS} = 15^{+5}_{-3} ~M_{\odot}$ \citep{leonard2}. 
Contrary to these, the progenitors of SN~2003gd \citep{vandyk1, smartt1}
and SN~2005cs \citep{li2, maund2}, also detected on pre-explosion images,
are found within the mass range of $M_{ZAMS} = 7 - 9 ~M_{\odot}$. These
two recent discoveries have led to a conclusion that perhaps all Type IIp
SNe emerge from relatively less massive progenitors, so that 
$M_{ZAMS} = 8 - 15 ~M_{\odot}$. This is supported by the non-detection
of the progenitors of SN~1999em, SN~1999gi and SN~2001du \citep{smartt}
giving $M_{ZAMS} < 15 ~M_{\odot}$. 

However, the possible masses of the progenitor of SN~2004dj determined above 
do not fully support this scenario. If the $T_{cl}=8$ Myr solution of S96 is indeed
valid, then the progenitor mass of SN~2004dj must have been higher 
than $20 ~M_{\odot}$. This is significantly higher than the masses found
for the other progenitors, even for SN~1999ev (see above). Since the other
two solutions for S96 give progenitor masses that are consistent with the
mass range of the others, it cannot be stated that SN~2004dj had definitely
a more massive progenitor, but it cannot be ruled out at present. 
In principle, the cluster masses in Table~13 imply that there might be
at least a few stars with $M > 20 ~M_{\odot}$ when SN~2004dj exploded. 
From the integration of the Salpeter IMF, the number of stars with 
$M > 20 ~M_{\odot}$ is $N(M>20) ~\approx~ 2 \cdot (M / 10^3 M_{\odot})$,
which gives $N(M>20) = 58$ for $M_{cl} = 29 \cdot 10^3 ~M_{\odot}$. 
The actual number of massive stars in S96 may, of course, differ from
this value, but statistically it is possible that at least a few such
stars are indeed there. \citet{maiz} estimated that S96 contains about
12 red and perhaps 2-3 blue supergiants, based on their SED parameters.
The spectrophotometric observations of the
remainder of S96 after the SN faded away would help to clarify the picture. 

Comparing the observations of the CO vibrational bands in the mid-IR
with theoretical explosion models combined with radiative transfer calculations,
\citet{kotak} presented evidence for a red supergiant with $\sim 15 ~M_{\odot}$
as the likely progenitor of SN~2004dj. Again, this is at the upper end of the 
detected progenitor masses for Type II SNe, close to the prediction of
the $\sim 14$ Myr cluster solution by \citet{maiz}. 

\citet{zhang} estimated the parameters of the pre-supernova star using the
formulae given by \citet{nady}. These formulae relate the length of the plateau
$\Delta t_p$, the middle-plateau absolute magnitude $M_V$ and expansion velocity
at the photosphere $v_{ph}$ to the explosion energy ($E_{exp}$), ejected mass
($M_{ej}$) and progenitor radius ($R$). 
\citet{zhang} derived $\Delta t_p = 80 \pm 21$ days, $M_V = -16.55 \pm 0.35$ mag and
$v_{ph} = 3933 \pm 189$ kms$^{-1}$, and they got
$E_{exp} = 0.75^{+0.56}_{-0.38} \times 10^{51}$ erg, 
$M_{ej} = 10^{+7}_{-5} ~M_\odot$ and $R = 282^{+253}_{-122} ~R_\odot$.
From the more detailed analysis presented in this paper, we estimate these
parameters as $\Delta t_p = 100 \pm 20$ days (Sect. 2.1.1, 
allowing $\delta t \approx 10$ days for the peak duration before the plateau, 
see \citet{nady}), $M_V = -15.88 \pm 0.18$ mag and $v_{ph} = 3250 \pm 250$ kms$^{-1}$
(Sect. 4.2 and 4.3). These parameters result in 
$E_{exp} \approx 0.86_{-0.49}^{+0.89} \times 10^{51}$ erg, 
$M_{ej} \approx 19_{-10}^{+20} ~ M_\odot$ and 
$R \approx 155_{-75}^{+150} ~R_\odot$. It is interesting that
the ejected mass is, again, close to $\sim 20 ~M_\odot$, although
with high uncertainty.  It is concluded that
the presently available information suggest that the progenitor mass of
SN~2004dj is $M_{prog} \geq 15 ~M_{\odot}$. The SED fitting of S96 may
suggest a cluster age as low as 8 Myr, indicating $M_{prog} > 20 M_\odot$.
This massive progenitor may also be suspected from theoretical SN models.

\section{Conclusions}

The conclusions of this paper are summarized as follows:

\begin{itemize}
\item{New $BVRI$ photometric observations of SN~2004dj are presented. 
The date of explosion is estimated as June 30, 2004 (JD $2453187 \pm 20$),
about 1 month before discovery. The plateau phase lasted 
about $\sim 110 \pm 20$ days after explosion. 
The most probable value of the reddening
toward SN~2004dj is $E(B-V) = 0.07 \pm 0.1$ mag \citep{gk}. 
The progenitor cluster S96 is also detected on pre-explosion frames.}

\item{The new, plateau-phase optical spectra reveal great similarity between 
SN~2004dj and SN~1999em, a typical SN IIp. Radial velocities from metallic
lines and $H\alpha$ are presented. The spectral features are successfully 
modeled with the spectrum synthesis code {\it SYNOW} with $T_{bb}=8000$ K
as the photospheric temperature. The observed P Cyg profile of 
$H\alpha$ can be better described by an increased emission of the 
photosphere relative to the blackbody intensity at the rest wavelength 
of $H\alpha$. A single nebular spectrum confirms the $H\alpha$ asymmetry 
reported by \citet{chugai}. }

\item{The distance to SN~2004dj is estimated from two methods applicable for 
SNe IIp. The EPM distance is $D_{EPM} = 3.66 \pm 0.3$ Mpc, while
the SCM gives $D_{SCM} = 3.50 \pm 0.2$ Mpc. These are in good agreement with
other distance estimates to the host galaxy NGC~2403, 
based on Cepheids and TF-relation, although the Cepheid-distance is 
slightly less than the SN-distances. 
The averaged distance is $D = 3.47 \pm 0.3$ Mpc, 
which is consistent with all available information.}

\item{Using the updated distance, the nickel mass of SN~2004dj is calculated as
$M_{Ni} = 0.02 \pm 0.01 ~M_{\odot}$ from the shape of the tail light curve. 
The progenitor mass has been estimated from fitting theoretical SEDs to
the observations of S96. The solutions are similar to those of \citet{maiz}
and \citet{wang}, except that a "young" solution corresponding to 
$T_{cl} = 8$ Myr is also possible. Such a young cluster age would mean
that the mass of the progenitor was $M > 20 ~M_{\odot}$. The SN-masses 
of the other solutions are less than $15 ~M_{\odot}$, being in better 
agreement with other SNe IIp progenitors. }

\end{itemize}

\section*{Acknowledgments}
This research has been supported by Hungarian OTKA Grants No.TS 049872 and T042509.
The authors are grateful to the directors and staff of Konkoly Observatory, Szeged
Observatory (Hungary), David Dunlap Observatory (Canada) and F. L. Whipple Observatory
(USA) for generously allocating telescope time. They also want to express their 
warm thanks to Prof. David Branch and his group members at University of Oklahoma 
for providing access and support to the SYNOW code. 
Thanks are also due to an anonymous referee for the very thorough report and numerous
suggestions that helped us to improve the paper.
The NASA Astrophysics Data System,
the SIMBAD and NED databases, the Canadian Astronomy Data Centre and the Supernova
Spectrum Archive (SUSPECT) were used to access data and references. The availability
of these services are gratefully acknowledged. 

\label{lastpage}


\begin{thebibliography}{1}


\bibitem[\protect\citeauthoryear{Adelman et al.}{1989}]{adelman}
Adelman S.J. et al., 1989, A\&AS 81, 221

\bibitem[\protect\citeauthoryear{Baron et al.}{1994}]{bhb}
Baron E., Hauschildt P.H., Branch D., 1994, ApJ 426, 334

\bibitem[\protect\citeauthoryear{Baron et al.}{2000}]{baron1}
Baron E. et al., 2000, ApJ 545, 444

\bibitem[\protect\citeauthoryear{Baron et al.}{2004}]{baron2} 
Baron E., Nugent P.E., Branch D., Hauschildt P.H., 2004, ApJ 616, L91

\bibitem[\protect\citeauthoryear{Beswick et al.}{2005}]{besw}
Beswick R.J. et al. 2005, ApJ 623, L21

\bibitem[\protect\citeauthoryear{Branch et al.}{2003}]{branch1}
Branch D. et al. 2003, AJ 126, 1489

\bibitem[\protect\citeauthoryear{Branch et al.}{2004}]{branch2}
Branch, D. et al. 2004, ApJ 606, 413

\bibitem[\protect\citeauthoryear{Bruzual \& Charlot}{2003}]{bc03}
Bruzual G, Charlot S., 2003, MNRAS 344, 1000

\bibitem[\protect\citeauthoryear{Chugai et al.}{2005}]{chugai}
Chugai N.N. et al. 2005, Ast.L. 31, 792

\bibitem[\protect\citeauthoryear{Dessart \& Hillier}{2005a}]{dh1}
Dessart L., Hillier D.J., 2005 A\&A 439, 617

\bibitem[\protect\citeauthoryear{Dessart \& Hillier}{2005b}]{dh2}
Dessart L., Hillier D.J., 2006, A\&A 447, 691

\bibitem[\protect\citeauthoryear{Eastman, Schmidt \& Kirshner}{1996}]{eastman}
Eastman R.G., Schmidt B.P., Kirshner R., 1996, ApJ 466, 911 

\bibitem[\protect\citeauthoryear{Elmhamdi et al.}{2003a}]{elmha1}
Elmhamdi A., Chugai N.N., Danziger I.J., 2003, A\&A 404, 1077

\bibitem[\protect\citeauthoryear{Elmhamdi et al.}{2003b}]{elmha2}
Elmhamdi A. et al. 2003, MNRAS 338, 939

\bibitem[\protect\citeauthoryear{Fisher}{2000}]{fisher}
Fisher, A. 2000, PhD Thesis, Univ. Oklahoma

\bibitem[\protect\citeauthoryear{Fraternali et al.}{2001}]{frater}
Fraternali F., Oosterloo T, Sancisi R., van Moorsel G., 2001, ApJ 562, L47

\bibitem[\protect\citeauthoryear{Freedman et al.}{2001}]{freedman1}
Freedman W.L., Madore B.F., Gibson B.K. et al., 2001, ApJ 553, 47

\bibitem[\protect\citeauthoryear{Girardi et al.}{1996}]{girardi}
Girardi L. et al., 1996, A\&AS 117, 113 

\bibitem[\protect\citeauthoryear{Guenther \& Klose}{2004}]{gk} 
Guenther E.W., Klose S., 2004, IAU Circ. No. 8384

\bibitem[\protect\citeauthoryear{Hamuy et al.}{2001}]{hamuy} 
Hamuy M. et al., 2001, ApJ, 558, 615 

\bibitem[\protect\citeauthoryear{Hamuy \& Pinto}{2002}]{hp} 
Hamuy M., Pinto P.A., 2002, ApJ 566, L63

\bibitem[\protect\citeauthoryear{Hamuy}{2005}]{hamuy1} 
Hamuy, M. 2005, in Marciade J.M., Weiler K.W. eds., Proc. IAU Colloq. 192: Cosmic Explosions, 
On the 10th Anniversary of SN1993J, Springer Proceedings in Physics, p.535

\bibitem[\protect\citeauthoryear{Hendry et al.}{2005}]{hendry}
Hendry, M.A., Smartt, S.J., Maund, J.R. et al., 2005, MNRAS 359, 906 

\bibitem[\protect\citeauthoryear{Jimenez et al.}{2004}]{jim}
Jimenez R. et al., 2004, MNRAS 349, 240

\bibitem[\protect\citeauthoryear{Kasen et al.}{2002}]{kasen}
Kasen D., Branch D., Baron E., Jeffery D., 2002, ApJ 565, 380

\bibitem[\protect\citeauthoryear{Kharitonov et al.}{1988}]{kharit}
Kharitonov A.V., Tereshchenko V.M., Kunyazeva L.N., 1988, in Spectrophotometric Catalogue of Stars
(Alma-Ata, Nauka) p. 484

\bibitem[\protect\citeauthoryear{Kirshner \& Kwan}{1974}]{kikw}
Kirshner R.P., Kwan J., 1974, ApJ 193, 27

\bibitem[\protect\citeauthoryear{Korcakova et al.}{2005}]{korcak}
Korc\'akov\'a D. et al., 2005, IBVS No. 5605

\bibitem[\protect\citeauthoryear{Kotak et al.}{2005}]{kotak}
Kotak R. et al., 2005, ApJ 628, L123

\bibitem[\protect\citeauthoryear{Larsen}{1999}]{larsen}
Larsen S.S., 1999 A\&AS 139, 393

\bibitem[\protect\citeauthoryear{Leonard et al.}{2002a}]{leonard1} 
Leonard D.C. et al., 2002, PASP 114, 35

\bibitem[\protect\citeauthoryear{Leonard et al.}{2002b}]{leonard2} 
Leonard D.C. et al., 2002, AJ 124, 2490

\bibitem[\protect\citeauthoryear{Leonard et al.}{2003}]{leonard3} 
Leonard D.C., Kanbur S.M., Ngeow C.C., Tanvir N.R., 2003, ApJ 594, 247

\bibitem[\protect\citeauthoryear{Li et al.}{2005a}]{li1}
Li W., Van Dyk S.D., Filippenko A.V., Cuillandre J.-C., 2005, PASP 117, 121 

\bibitem[\protect\citeauthoryear{Li et al.}{2005b}]{li2}
Li W. et al. 2005, preprint (astro-ph/0507394)

\bibitem[\protect\citeauthoryear{Ma\'iz-Apell\'aniz et al.}{2004}]{maiz}
Ma\'iz-Apell\'aniz J. et al., 2004, ApJ 615, L113

\bibitem[\protect\citeauthoryear{Maund \& Smartt}{2005}]{maund1}
Maund J.R., Smartt S.J., 2005, MNRAS 360, 288

\bibitem[\protect\citeauthoryear{Maund, Smartt \& Danziger}{2005}]{maund2}
Maund J.R., Smartt S.J., Danziger I.J., 2005, MNRAS 364, 33

\bibitem[\protect\citeauthoryear{Nadyozhin}{2003}]{nady} 
Nadyozhin D.K. 2003 MNRAS 346, 97

\bibitem[\protect\citeauthoryear{Nakano et al.}{2004}]{nakano} 
Nakano, S. et al. , 2004, IAU Circ. No. 8377

\bibitem[\protect\citeauthoryear{Patat et al.}{2004}]{patat1}
Patat F., Benetti S., Pastorello A., Filippenko A.V., 2004, IAU Circ. No. 8378

\bibitem[\protect\citeauthoryear{Phillips \& Williams}{1991}]{pw}
Phillips M.M., Williams R.E. 1991 in Woosley S.E. ed. Supernovae, Springer, N.Y., p. 36.

\bibitem[\protect\citeauthoryear{Pilyugin, Vilchez \& Contini}{2004}]{pily}
Pilyugin L.S., Vilchez J.M., Contini T., 2004, A\&A 425, 849

\bibitem[\protect\citeauthoryear{Pooley \& Lewin}{2004}]{pool}
Pooley D., Lewin W.H.G., 2004, IAU Circ. 8390

\bibitem[\protect\citeauthoryear{Riess et al.}{2005}]{riess}
Riess A.G. et al., 2005, ApJ 627, 579

\bibitem[\protect\citeauthoryear{Russell}{2002}]{russell}
Russell D.G., 2002, ApJ 565, 681

\bibitem[\protect\citeauthoryear{S\'arneczky et al.}{2005}]{sry}
S\'arneczky K. et al., 2005 in Marciade J.M., 
Weiler K.W. eds., Proc. IAU Colloq. 192: Cosmic Explosions, 
On the 10th Anniversary of SN1993J, Springer Proceedings in Physics, CD-ROM

\bibitem[\protect\citeauthoryear{Schlegel et al.}{1998}]{sfd}
Schlegel D., Finkbeiner D., Davis M., 1998, ApJ 500, 525

\bibitem[\protect\citeauthoryear{Sckrutskie et al.}{1997}]{skrut}
Skrutskie M.F. et al 1997 in Garz\'on F. et al. eds, 
The Impact of Large Scale Near-IR Sky Surveys, ASSL Vol.210, Kluwer, p.25  

\bibitem[\protect\citeauthoryear{Smartt et al.}{2003}]{smartt}
Smartt S.J. et al., 2003, MNRAS 343, 735

\bibitem[\protect\citeauthoryear{Smartt et al.}{2004}]{smartt1}
Smartt S.J. et al., 2004, Science 303, 499

\bibitem[\protect\citeauthoryear{Stockdale et al.}{2004}]{stock}
Stockdale, C.J. et al., 2004, IAU Circ. 8379

\bibitem[\protect\citeauthoryear{Sugerman \& Van Dyk}{2005}]{suger}
Sugerman B., Van Dyk S., 2005, IAU Circ. No. 8489

\bibitem[\protect\citeauthoryear{Suntzeff}{2000}]{suntzeff}
Suntzeff N.B., 2000, American Institute of Physics Conference Series, 522, 65 

\bibitem[\protect\citeauthoryear{Van Dyk et al.}{2003}]{vandyk1} 
Van Dyk S.D., Li W., Filippenko A.V., 2003 PASP 115, 1289

\bibitem[\protect\citeauthoryear{Vink\'o et al.}{2004}]{vinko1}
Vink\'o, J. et al., 2004, A\&A 427, 453 

\bibitem[\protect\citeauthoryear{Wang et al.}{2005}]{wang} 
Wang, X. et al., 2005, ApJ 626, 89

\bibitem[\protect\citeauthoryear{Woosley \& Weaver}{1986}]{ww} 
Woosley S.E., Weaver T.A., 1986, Ann. Rev. Astron. Astroph. 24, 205

\bibitem[\protect\citeauthoryear{Yamaoka et al.}{2004}]{yama}
Yamaoka K., Ma\'iz-Apell\'aniz J., Bond H.E., Siegel M.H., 2004, IAU Circ. No. 8385

\bibitem[\protect\citeauthoryear{Zhang et al.}{2005}]{zhang} 
Zhang, T. et al., 2005, preprint (astro-ph/0512526)


\end{thebibliography}
\end{document}